\documentclass[floatfix,%
 reprint,
nofootinbib,
 amsmath,amssymb,
 aps,
prb,
]{revtex4-2}
\usepackage{subfigure}
\usepackage{graphicx}
\usepackage{dcolumn}
\usepackage{bm}
\usepackage[normalem]{ulem}
\usepackage{hyperref}
\usepackage{xcolor}

\begin{document}

\title{Helical phases and Bogoliubov Fermi surfaces probed by superconducting diode effects}
\author{Zekun Zhuang}
\author{Daniel Shaffer}
\author{Jaglul Hasan}
\author{Alex Levchenko}
\affiliation{Department of Physics, University of Wisconsin-Madison, Madison, Wisconsin 53706, USA}

\date{\today}

\begin{abstract}
Noncentrosymmetric superconductors (NCSs) with Rashba spin–orbit coupling (SOC) and in-plane magnetic fields have emerged as natural platforms for realizing both the bulk superconducting diode effect (SDE) and the Josephson diode effect (JDE) -- phenomena characterized by unequal critical currents in opposite directions due to the simultaneous breaking of time-reversal and inversion symmetries.
Using the quasiclassical Eilenberger formalism, we systematically investigate both the bulk SDE and the JDE in a clean NCS with Rashba SOC and in-plane magnetic fields. For the bulk system, we find that the diode efficiency can nominally approach its maximal value at the critical endpoint of the first-order Lifshitz transition between weak and strong helical phases featuring finite-momentum Cooper pairs, the latter marked by the emergence of Bogolyubov Fermi surfaces (BFSs).
In a Josephson junction, we show that finite-momentum pairing in the superconducting leads is the dominant mechanism behind the JDE in short junctions, whereas in long junctions it is primarily governed by the Zeeman field in the normal region. In the long-junction regime, the diode efficiency additionally oscillates between positive and negative values as a function of magnetic field at low fields, providing a route toward a highly tunable Josephson diode.
At higher fields, the onset of BFSs in the strong helical phase leads to a sharp suppression of both the JDE and the Josephson current when the current direction is aligned with momenta along the BFS, resulting in strong anisotropy. We propose that this anisotropy in the Josephson current offers an alternative method for detecting BFSs, applicable to systems with or without a JDE.
\end{abstract}

\maketitle

\section{Introduction}

There has recently been a resurgence of interest in noncentrosymmetric superconductors (NCS) \cite{BauerSigrist12, SmidmanAgterberg17} lacking inversion symmetry due to their potential applications for realizing the bulk superconducting diode effect (SDE), the anomalous Josephson effect (AJE), and the Josephson diode effect (JDE) \cite{Nadeem23, NagaosaYanase24, Ma25}. 
Both SDE and JDE -- defined as the inequality of the critical currents \(|I_{c+}|\neq |I_{c-}|\) flowing in opposite directions -- require the breaking of inversion symmetry, as well as time-reversal symmetry (TRS). In 2D NCSs the former is broken by spin-orbit coupling (SOC) like Rashba SOC, while the latter is typically realized by applying in-plane magnetic fields, leading to the formation of helical SC states with finite-momentum Cooper pairing.
However, while helical superconductors have been investigated theoretically for a long time \cite{MineevSamokhin94, GorkovRashba01, BarzykinGorkov02, FrigeriAgterbergSigrist04, BauerSigrist12, SmidmanAgterberg17, DimitrovaFeigelman03, AgterbergKaur07, DimitrovaFeigelman07}, including in the context of bulk SDE \cite{LevitovNazarovEliashberg85, Edelstein96, DaidoYanase22, YuanFu22, IlicBergeret22, HasanShafferKhodasLevchenko24, HasanShafferKhodasLevchenko25, BankierLevchenkoKhodas25}, the experimental evidence for such states had been limited \cite{Chen18, HartYacoby17, MichaeliPotterLee12}. More broadly, the same is true for other nonuniform superconducting states such as FFLO \cite{FF, LO} and pair density waves \cite{Agterberg20}. Other exotic effects have been predicted in NCSs, including the superconducting Edelstein effect \cite{Edelstein95} and a transition at higher fields from a weak to a strong helical phase with a larger pairing momentum \cite{DimitrovaFeigelman03}, but those have also not been observed to the best of our knowledge. The strong helical phase is further characterized by a gapless spectrum due to the presence of Bogolyubov Fermi surfaces (BFS) \cite{DimitrovaFeigelman07, AgterbergKaur07, TimmAgterberg17, YuanFu18, LinkBoettcherHerbut20, ShafferBurnellFernandes20, AkbariThalmeier22, BanerjeeSchnyder22, BabkinSerbyn24, Mateos24, Sano25, Wei25}, which has also not been definitevely observed in NCSs.

Though in principle it is possible for a system to exhibit SDE without finite-momentum pairing\footnote{Without fine-tuning, this can happen due to symmetry, for example in systems with \(C_3\) three-fold rotation symmetry like that  considered in \cite{Zhai22}.}, recent observations of intrinsic bulk SDE in systems with Rashba SOC
\cite{AndoYanaseOno20, LinScheurerLi22, BauriedlParadiso22, KealhoferBalentsStemmer23, GengBergeretHeikkila23, YunKim23, LeLin24, AsabaYanaseMatsuda24, JamesLeo2024NatMat, LiuIwasa24, WanDuan24, Ingla-AynesMoodera24}
thus provide the strongest evidence of non-uniform helical superconductivity to date, assuming extrinsic effects can ultimately be ruled out \cite{HouMoodera23, MollGeshkenbein23}.
On the other hand, it has also been shown that finite-momentum pairing can lead to JDE as well \cite{DavydovaFu22, KochanZutic23, Zhang24, HuangVayrynen24}. Surprisingly, despite early proposals to use Josephson junctions between finite-momentum and uniforms SCs to identify helical phases in NCSs \cite{YangAgterberg00, KaurAgterbergSigrist05} as well as to realize the AJE \cite{Buzdin08, HasanSongciLevchenko22}, JDE with helical NCSs have not been considered either in experiment or in theory with the exception of \cite{MeyerHouzet24, Roig24, ZazunovEgger24} 
and, for the case of a diffusive SIS'IS junction, \cite{OsinLevchenkoKhodas24}. 
Instead, the vast majority of theoretical studies of JDE junctions assume uniform superconductivity, with the breaking of time-reversal and inversion symmetries by magnetic fields and Rashba SOC confined to the normal region of the junction
\cite{ReynosoAvignon08, ZazunovEggerMartin09, ReynosoAvignon12, YokoyamaNazarov13, YokoyamaNazarov14, NesterovHouzetMeyer16,PekertenMatosAbiague22, WangWangWu22, ZhangJiang22, ChengSun23, CostaKochan23, CostaManfraKochanParadisoStrunk23, Huang23,  LiuAndreevSpivak24,   PekertenMatosAbiagueZutic24, MeyerHouzet24, WangChen24, Soori24, VakiliKovalev24, IlicBergeret24, PatilBelzig24, CostaFabian24} (surface states of 3D TIs that can be modeled with Rashba SOC have also been considered \cite{TanakaLuNagaosa22, LuTanakaNagaosa23, LiuAndreevSpivak24, Karabassov24, KarabassovBobkovaVasenko24}; there are also alternative proposed mechanisms that do not rely on Rashba SOC \cite{DolciniMeyerHouzet15, ZinklSigrist22, WeiLiu22, ShafferLiLevchenko25}).
The same is generally true of experimental observations of JDE \cite{GoldmanKreisman67, KrasnovPedersen97, MendittoGoldobin16, DartiailhZutic21, BaumgartnerManfra22, TuriniGiazotto22, Ghosh24, Nadeem23, Ma25}, where either the normal region or extrinsic geometric effects are responsible for the effect. To the best of our knowledge, evidence of finite momentum pairing in JDE experiments has only been claimed in \cite{PalDavydovaFuParkin22}, but the finite momentum was attributed to screening currents or proximity effects in the normal region rather than a bulk property of the superconductor (JDE was also likely seen but not reported in \cite{Chen18}).

Motivated by these considerations, in this work we revisit the helical Rashba model and study both the SDE and JDE with the goal of identifying signatures of the helical states. While our focus is on JDE since SDE has been extensively investigated in this model, we note that the focus of most of the SDE studies had been on the low-field weak helical phase. The SDE in the strong helical phase, on the other hand, had only been studied in \cite{DaidoYanase22, Daido2022_2, IlicBergeret22}, who noted in particular that the superconducting diode efficiency \(\eta=(I_{c+}+I_{c-})/(I_{c+}-I_{c-})\) is enhanced in the vicinity of the transition before it then changes sign. This occurs both at a first-order phase transition that takes place at lower temperatures \cite{DaidoYanase22, Daido2022_2} and in the crossover regime at higher temperatures \cite{Daido2022_2, IlicBergeret22}.
Building on these results, we observe that formally the diode coefficient approaches its nominally perfect value of \(\pm 1\) precisely at the critical end point at which the first-order phase transition terminates, similar to the case of a tricritical point of the FFLO phase in conventional superconductors \cite{YuanFu22} and second-order phase transitions between uniform and non-uniform SCs \cite{ShafferChichinadzeLevchenko24, ChakrabortyBlackSchaffer24}.
We note that this result depends on the precise definition of \(I_{c\pm}\), as \(I(q)\), the supercurrent at fixed Cooper pair momentum \(q\), exhibits multiple local maxima and minima due to the presence of multiple local minima of the free energy. We anticipate that effects due to phase coexistence and temperature fluctuations reduce the diode efficiency, but we nevertheless expect that tuning the system close to the critical end point maximizes \(\eta\). This result illustrates that bulk SDE is often enhanced in the vicinity of critical points, suggesting a general guiding principle for maximizing the efficiency of bulk superconducting diodes.

We find that the signatures of helical phases are even more pronounced in the JDE, which we study by solving the Eilenberger equations and accounting for the non-uniform pairing in the SC regions self-consistently (which is strictly speaking necessary to preserve charge current conservation \cite{Krekels25}).
In the short junction limit, we find that the field-induced pairing momentum leads to JDE, which agrees with earlier results obtained in Ref. \cite{DavydovaFu22} where the pairing momentum is induced by the external screening current.
In the less-studied long junction limit, we find that the diode coefficient changes sign periodically as a function of the applied magnetic field when the superconductor is in the weak helical phase with period \(\sim E_T\pi/2\) where \(E_T=v_F/L\) is the Thouless energy, \(v_F\) is the Fermi velocity, and \(L\) is the length of the junction. When the junction length is finite, the periodicity originates from the relative phase shift between two helical bands, attributed to both the finite-momentum pairing in the superconductor and the Zeeman field in the normal metal, the latter of which dominates in a long junction.
We note that this can be useful for creating tunable and reversible Josephson diodes.
In the strong helical phase, in contrast, accounting for the Cooper pair momentum in the superconducting regions changes the JDE qualitatively: both the diode coefficient and the critical current itself are strongly suppressed.

We attribute this suppression, one of the main results of our work, to another important property of the strong helical phase, namely its gapless nature that leads to the formation of the BFS. 
The suppression is moreover strongly correlated with the current direction relative to the location of the BFS in momentum space: it is strongest when the Josephson current direction intersects with BFSs, and weaker if it does not, resulting in a strongly anisotropic response.
BFSs were predicted a long time ago to exist on general and topological grounds \cite{Volovik89, YangSondhi98, KarchevLittlewood01, LiuWilczek03, MatsuuraSchnyderRyu13, KobayashiSato14}, as well as due to the Volovik effect \cite{Volovik93, MolerKapitulnik94} and in nonuniform SC states including pair density waves \cite{BaruchOrgad08, BergFradkinKivelsonTranquada09, Gao10, SotoGarridoFradkin15,  Agterberg20}, but the term was first introduced in the context of inversion symmetric SCs with broken time-reversal symmetry and topologically protected BFSs \cite{Agterberg17, BzdusekSigrist17, BrydonAgterberg18, MenkeBrydon19, SumitaYanase19, LappTimm20, SumitaYanase19, SettyHirschfeld20_II, DuttaBlackSchaffer21, ShafferWangSantos21, PalDutta24}. It has subsequently been shown that such BFSs are generally unstable towards interactions that spontaneously break the inversion symmetry \cite{TamuraHoshino20, Oh20, OhAgterberg21, TimmAgterberg21, HerbutLink21, Mori25}, so the focus shifted again to realizations in NCSs. More recently BFSs have also been considered in (inversion-symmetric) altermagnets \cite{Wei24, deCarvalho24, HongPark25, WuWangFernandes25}.

A big challenge in identifying BFSs is the fact that the most accessible experimental observables are tied to the residual density of states (DOS), including heat capacity and thermal transport \cite{Volovik89, YangSondhi98, KarchevLittlewood01, LappTimm20, Mateos24, Pal24, Sano25}, penetration depth \cite{LappTimm20}, NMR relaxation rates \cite{LappTimm20, CaoHirschfeld24, Yu25}, optical conductivity \cite{OhAgterberg21}, and tunneling DOS and anomalous Fano factors \cite{Agterberg17, LappTimm20, BanerjeeSchnyder22, PalDutta24, Wei24, Ohashi24}; a finite zero-bias tunneling DOS is also known to exist when BFSs appear due to external currents \cite{Fulde65, DavydovaFu24}. A residual DOS, however, is also well-known to exist in SCs with magnetic disorder \cite{AG60, ReifWoolf62}, which means that such observables provide only indirect evidence of BFSs. Some effort has thus gone into identifying effects of disorder on BFSs \cite{Fulde65, OhAgterberg21, BabkinSerbyn24, MikiHoshino24}, but QPI and ARPES are otherwise the only direct measurements of BFSs that have been suggested in the literature \cite{Agterberg17, AkbariThalmeier22}. Topologically-protected BFSs are also expected to host topological surface states \cite{ShafferBurnellFernandes20, Mo25, RuizStrunk25}, but their detection would be challenging due to possible hybridization with gapless bulk states.

The strongly anisotropic suppression of JDE and the Josephson current in the presence of the BFS is not likely to be mimicked by disorder-induced gapless superconductivity, and therefore provides an alternative experimental probe for detecting BFSs that is not simply tied to the residual DOS.
Moreover, we posit that the anisotropic suppression of the Josephson current would generically occur in any multiband model with BFSs, regardless of the presence of JDE.
Note that in earlier studies, Josephson junctions with BFSs have only been considered in the short junction limit \cite{MikiHoshino24, CohenKhodas24, RuizStrunk25}, where no drastic effect is seen due to the BFS.
Our results therefore establish long Josephson junctions as a platform for probing both helical phases in NCSs and BFSs more generally.

The paper is organized as follows. In Sec. \ref{Sec:theory}, we introduce the Hamiltonian and solve the corresponding quasiclassical Eilenberger equation. We then present results for the bulk superconductor and SNS junction in Sec. \ref{Sec:result}, focusing on the properties of Andreev spectrum, current-phase relations and diode efficiency. We end by discussing our results in Sec. \ref{Sec:discussion}.

\section{Theoretical formulation}\label{Sec:theory}
\subsection{Model Hamiltonian}
\begin{figure}[t]
    \centering\subfigure[]{\includegraphics[width=0.48\textwidth]{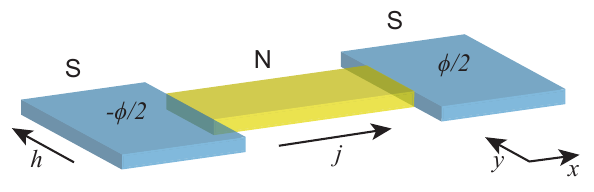}\label{fig:sch}}
    \centering\subfigure[]{\includegraphics[width=0.48\textwidth]{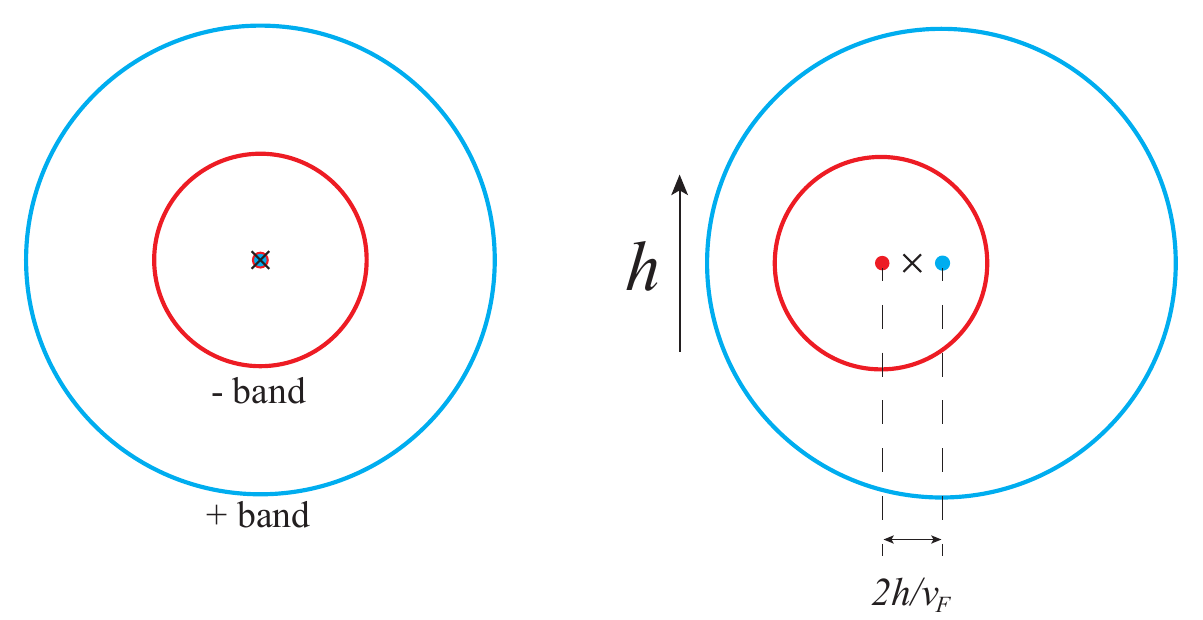}\label{fig:FS}}
    \caption{(a) The schematic setup of the S-N-S junction, where a helical metal is sandwiched between two helical superconductors; a homogeneous in-plane magnetic field $h$ is applied throughout the junction. (b) The Fermi surface of the normal metal when the magnetic field is absent (left) and present (right).}
    \label{fig:Schematic}
\end{figure}
We consider a strongly spin-orbit coupled superconducting SNS junction under an in-plane magnetic field in the ballistic limit, as shown in Fig. \ref{fig:sch}. The 2-D bulk normal metal is modeled by the Hamiltonian
\begin{equation}
    H=H_0+H_Z,
\end{equation}
where $H_0$ describes the 2D electron gas with strong Rashba SOC 
\begin{equation}    H_0=\sum_{\mathbf{k}} f_\mathbf{k}^\dagger\left[\epsilon_\mathbf{k}-\alpha(\sigma_x k_y-\sigma_y k_x)\right]f_\mathbf{k}, \label{eq:H0}
\end{equation}
in which $\epsilon_\mathbf{k}=k^2/2m-\mu$, $\mu$ is the chemical potential, $\alpha$ is the Rashba spin-orbit coupling strength and we take shorthand $f_\mathbf{k}\equiv(f_{\mathbf{k}\uparrow},f_{\mathbf{k}\downarrow})^T$.  The Zeeman term $H_Z$ is due to the in-plane magnetic field
\begin{equation}
    H_Z=\sum _{\mathbf{k}} f_\mathbf{k}^\dagger \mathbf{h}\cdot\boldsymbol{\sigma} f_\mathbf{k}. \label{eq:HZ}
\end{equation}
In the 2-D superconductor, an additional term $H_S$ arises,
\begin{equation}
    H_S=\frac{1}{2}\sum_{\mathbf{k}} \Delta_\mathbf{q}(f_{\mathbf{k}\uparrow}^\dagger f_{-\mathbf{k+q}\downarrow}^\dagger-f_{\mathbf{k}\downarrow}^\dagger f_{-\mathbf{k+q}\uparrow}^\dagger)+\text{H.c.},\label{eq:HS}
\end{equation}
which is due to the singlet pairing of electrons with total momentum $\mathbf{q}$. Throughout this work we set $\hbar=k_B=1$ and assume $\alpha>0$ without loss of generality.

The Hamiltonian $H_0$ can be diagonalized in the helical band basis $\lambda=\pm$ by making the unitary transformation  $f_\mathbf{k}=U_\mathbf{k} c_\mathbf{k}$, $c_\mathbf{k}\equiv(c_{\mathbf{k}+},c_{\mathbf{k}-})^T$, $U_\mathbf{k}=\frac{1}{\sqrt{2}}\left(\begin{array}{cc}
    1 & 1 \\
    -ie^{i\phi_\mathbf{k}}& ie^{i\phi_\mathbf{k}}  
\end{array}\right)$, where $\phi_\mathbf{k}$ is the angle between $\mathbf{k}$ and $x$-axis. In this basis the Hamiltonian becomes
\begin{equation}
H_0=\sum_{\mathbf{k},\lambda}\left(\epsilon_\mathbf{k}-\lambda \alpha k\right)c_{\mathbf{k}\lambda}^\dagger c_{\mathbf{k}\lambda},
\end{equation}
\begin{equation}
    H_S=\frac{1}{2}\sum_{\lambda,\mathbf{k}}i\lambda e^{-i \phi_\mathbf{k}}\Delta_q c_{\mathbf{k} \lambda}^\dagger c_{-\mathbf{k+q} \lambda}^\dagger+\text{H.c}.,
\end{equation}
\begin{equation}
    H_Z=\sum_{\mathbf{k},\lambda}c_\mathbf{k}^\dagger \mathbf{h}_\mathbf{k}\cdot \boldsymbol{\sigma}c_\mathbf{k},\label{eq:HZ2}
\end{equation}
where $\mathbf{h}_\mathbf{k}=(0,-\mathbf{h}\cdot\hat{\mathbf{k}},-(\mathbf{h} \times \hat{\mathbf{z}})\cdot\hat{\mathbf{k}})$, $\hat{\mathbf{k}}=\mathbf{k}/k$. At zero magentic field, the two helical bands have equal Fermi velocity $v_\lambda=v_F=\sqrt{2\mu/m+\alpha^2}$, and different density of states (DOS) $N_\lambda=N_0(1+\lambda\tilde{\alpha})$ where $N_0=m/2\pi$ and $\tilde{\alpha}=\alpha/v_F$.
In this work we consider weak Zeeman coupling $H_Z$, the lowest-order effect of which is to shift the center of the Fermi surface of  band $\lambda$ to a nonzero momentum $\mathbf{Q}_\lambda=\lambda(\mathbf{h}\times \hat{\mathbf{z}})/v_F$, as shown in Fig. \ref{fig:FS} . 

\subsection{Quasiclassical Eilenberger equation}
In the quasiclassical approximation, one can derive the Eilenberger equation by integrating the microscopic Green's function over its momentum amplitude\cite{Belzig1999,Bennemann2008,Eilenberger1968}. In the strong spin-orbit coupling limit and under a weak magnetic field, the Eilenberger equation of two helical bands decouples, and takes the simple form in the helical basis \cite{AgterbergKaur07,Houzet2015,IlicBergeret22}
\begin{equation}
    -i v_F \hat{\mathbf{k}}\cdot \boldsymbol{\nabla}_\lambda g_\lambda=[(i\omega_n-\hat{\Delta})\tau_z,g_\lambda], \label{eq:Eil}
\end{equation}
where $g_\lambda(\mathbf{r},\hat{\mathbf{k}},i\omega_n)$ is the 2-by-2 quasiclassical Green's function in the Nambu basis , $\boldsymbol{\nabla}_\lambda=\boldsymbol{\nabla}-i \mathbf{Q}_\lambda [\tau_z,...]$, $\omega_n=(2n+1)\pi T$ is the Matsubara frequency and
\begin{equation}
        \hat{\Delta}=\left(\begin{array}{cc}
       0  & \Delta(\mathbf{r}) \\
       \Delta^*(\mathbf{r})  & 0
    \end{array}\right).
\end{equation}
Note that we have ignored the higher-order effects of the magnetic field, e.g. due to the interband pairing and deformed Fermi surface shape \cite{HasanShafferKhodasLevchenko24,HasanShafferKhodasLevchenko25}, which are negligible in the regime $\mu,\tilde{\alpha}\mu\gg \Delta_q,h$ considered in this work. The self-consistent condition can be written in terms of the zero-field critical temperature $T_c$
\begin{equation}
    \Delta \ln \frac{T}{T_c}+\pi T\sum_{\omega_n>0,\lambda}\left[\frac{\Delta}{\omega_n}-(1+\lambda\tilde{\alpha})\langle  i(g_\lambda)_{12}\rangle_{\text{FS}}\right]=0 \label{SCDelta2}
\end{equation}
The local density of states (LDOS) of band $\lambda$ is given by
\begin{equation}
    \nu_\lambda(\omega)=\frac{N_\lambda}{2} \left\langle \text{Re}\{\text{Tr}[\tau_z g_\lambda(i\omega_n\rightarrow \omega+i\delta)]\}\right\rangle_{\text{FS}}, \label{LDOS}
\end{equation}
where $\langle ...\rangle_{_{\text{FS}}}$ denotes the angular average over the Fermi surface. The particle current density contributed by band $\lambda$ is given by
\begin{equation}
    \mathbf{j}_\lambda=-\frac{i\pi N_\lambda v_F T}{2}\sum_{\omega_n}\langle\text{Tr}[\tau_z g_\lambda\hat{\mathbf{k}}]\rangle_{\text{FS}}. \label{current}
\end{equation}

\subsubsection{Bulk superconductor}
To solve the Eilenberger equation, it proves convenient to make gauge transformation
\begin{equation}
    \Tilde{g}_\lambda=e^{-i\tau_z \mathbf{Q}_\lambda\cdot \mathbf{r}}g_\lambda e^{i\tau_z \mathbf{Q}_\lambda\cdot \mathbf{r}}, \label{eq:gaugeTrans}
\end{equation}
which transforms Eq. (\ref{eq:Eil}) to
\begin{equation}
        -i v_F \hat{\mathbf{k}}\cdot \boldsymbol{\nabla} \tilde{g}_\lambda=[(i\omega_n-\hat{\Delta}_\lambda)\tau_z,\tilde{g}_\lambda], \label{Eil3}
\end{equation}
where 
\begin{equation}    
\hat{\Delta}_\lambda=e^{-i\tau_z \mathbf{Q}_\lambda\cdot \mathbf{r}} \hat{\Delta} e^{i\tau_z \mathbf{Q}_\lambda\cdot \mathbf{r}}\\=\left(\begin{array}{cc}
       0  & \Delta_\lambda(r) \\
       \Delta_\lambda^*(r) & 0
    \end{array}\right),
\end{equation}
where $\Delta_\lambda(\mathbf{r})=\Delta_qe^{i\mathbf{q}_\lambda\cdot \mathbf{r}}$, $\mathbf{q}_\lambda\equiv\mathbf{q}-2\mathbf{Q}_\lambda$.
The homogeneous solution is given by
\begin{equation}
    \tilde{g}_\lambda^s(r)=\frac{1}{\Omega_\lambda}\left(\begin{array}{cc}
       \tilde{\omega}_{n,\lambda}  & -i\Delta_\lambda(\mathbf{r})\\
      i\Delta_\lambda^*(\mathbf{r})   & -\tilde{\omega}_{n,\lambda}
    \end{array}\right) \label{eq:homoSolution},
\end{equation}
where $\tilde{\omega}_{n,\lambda}=\omega_n+i v_F q_\lambda/2$, $\Omega_\lambda=\sqrt{|\Delta_\lambda|^2+\tilde{\omega}_{n,\lambda}^2}$ and the amplitude of $\Delta_q$ can be obtained by solving Eq. (\ref{SCDelta2}) self-consistently. At equilibrium, the value of $q=|\mathbf{q}|$ is determined by the requirement of vanishing supercurrent (\ref{current}).

\subsubsection{SNS junction}
Now we consider the SNS junction case with phase difference $\phi$. We neglect the inverse proximity effect and assume 
\begin{equation}
    \Delta(\mathbf{r})=\left\{
    \begin{array}{ll}
        \Delta_q e^{i[\mathbf{q}\cdot (\mathbf{r}+\frac{L}{2}\hat{\mathbf{x}}) -\phi/2]}, & x<-L/2  \\
        \Delta_q e^{i[\mathbf{q}\cdot (\mathbf{r}-\frac{L}{2}\hat{\mathbf{x}}) +\phi/2]}, & x>L/2 \\
        0, & |x|<L/2 \\
    \end{array}\right.
\end{equation}
where $\Delta_q>0$. The value of $q$ and $\Delta_q$ is determined numerically for the bulk superconductor at equilibrium (see Sec. \ref{sec:bulk}). As in the bulk case, one can make a gauge transformation (\ref{eq:gaugeTrans}) which absorbs the contribution of the magnetic field into the modulation vector of the pairing potential
\begin{equation}
    \Delta_\lambda(\mathbf{r})=\left\{
    \begin{array}{ll}
        \Delta_q e^{i[\mathbf{q}_\lambda\cdot (\mathbf{r}+\frac{L}{2}\hat{\mathbf{x}}) -\phi_\lambda/2]}, & x<-L/2  \\
        \Delta_q e^{i[\mathbf{q}_\lambda\cdot (\mathbf{r}-\frac{L}{2}\hat{\mathbf{x}}) +\phi_\lambda/2]}, & x>L/2 \\
        0, & |x|<L/2 \\
    \end{array}\right. \label{transformedDelta}
\end{equation}
where $\phi_\lambda=\phi-2Q_{\lambda x} L$. 

The current of each helical band can be found by solving the Eilenberger equation along each trajectory, assuming that the quasiclassical Green's function is continuous at $x=\pm L/2$. We first focus on the trajectory along the positive $x$-direction, i.e. $\hat{k}=(1,0)$. Inside the superconductor region, besides the homogenous solution (\ref{eq:homoSolution}) we have shown, there exist two other evanescent solutions \cite{Nikolic2019}
\begin{multline}
    \tilde{g}_{\lambda,\pm}^s(\mathbf{r})=\frac{e^{\pm 2\Omega_\lambda x/v_F}}{2\Omega_\lambda}\times\\\left(\begin{array}{cc}
       \Delta_q  & i(\tilde{\omega}_{n,\lambda} \mp\Omega_\lambda)\frac{\Delta_\lambda(\mathbf{r})}{\Delta_q}\\
      -i(\tilde{\omega}_{n,\lambda} \pm\Omega_\lambda)\frac{\Delta^*_\lambda(\mathbf{r})}{\Delta_q}   & -\Delta_q
    \end{array}\right).
\end{multline}
The solution in the normal region can be regarded as the special case by taking $\Delta_q=0$, which leads to
\begin{equation}
    \tilde{g}_{\lambda}^n(\mathbf{r})=\text{sgn}(\omega_n)\tau_z,
\end{equation}
\begin{equation}
    \tilde{g}_{\lambda,\pm}^n(\mathbf{r})=\tau_{\pm}e^{\mp 2\omega_n/v_F x},
\end{equation}
where $\tau_{\pm}=(\tau_x\pm i\tau_y)/2$. In each region the quasiclassical Green's functions are linear superposition of these solutions, namely
\begin{equation}
    \tilde{g}_\lambda(\mathbf{r})=\left\{\begin{array}{ll}
\tilde{g}_{\lambda}^s(\mathbf{r})+B^{\lambda}_{1}\tilde{g}_{\lambda,+}^s(\mathbf{r}),& x<-L/2  \\
         A^\lambda\tilde{g}_\lambda^n(\mathbf{r})+A^{\lambda}_{1}\tilde{g}_{\lambda,-}^n(\mathbf{r})+A^{\lambda}_{2}\tilde{g}_{\lambda,+}^n(r),& |x|<L/2\\
         \tilde{g}_\lambda^s(\mathbf{r})+B^{\lambda}_{2}\tilde{g}_{\lambda,-}^s(\mathbf{r}),& x>L/2  \\
    \end{array}\right..
\end{equation}
Assuming perfect interfaces at which the quasiclassical Green's functions are continuous, one can show that the coefficients are
\begin{equation}
    A^\lambda=-i\frac{i\tilde{\omega}_{n,\lambda}+\Omega_\lambda \tan\Phi_{\lambda,n}}{\Omega_\lambda-i\tilde{\omega}_{n,\lambda}\tan\Phi_{\lambda,n}},
\end{equation}
\begin{equation}
    A^\lambda_{1}=-A^{\lambda}_{2}=\frac{\Delta}{\Omega_\lambda \cos\Phi_{\lambda,n}-i\tilde{\omega}_{n,\lambda}\sin\Phi_{\lambda,n}},
\end{equation}
\begin{equation}
    B^{\lambda}_{1}=B^{\lambda}_{2}=\frac{-2i\Delta e^{\Omega_\lambda L/v_F}}{\Omega_\lambda \cot\Phi_{\lambda,n}-i\tilde{\omega}_{n,\lambda}}.
\end{equation}
Here $\Phi_{\lambda,n}=i\omega_n L/v_F-\phi_\lambda/2$. For a general trajectory $\hat{\mathbf{k}}^\pm=(\pm\cos\theta,\sin\theta)$, $-\pi/2<\theta<\pi/2$, the coefficient $A$ is given by
\begin{equation}
    A^\lambda_\pm=-i\frac{i\tilde{\omega}^\pm_{n,\lambda}+\Omega^\pm_\lambda \tan\Phi_{\lambda,n}^\pm}{\Omega^\pm_\lambda-i\tilde{\omega}_{n,\lambda}^\pm\tan\Phi_{\lambda,n}^\pm}, \label{Aeta}
\end{equation}
where $\tilde{\omega}_{n,\lambda}^\pm=\omega_n+ i(\mathbf{q}_\lambda\cdot\hat{\mathbf{k}}^\pm)v_F/2$, $\Omega_\lambda^\pm=\sqrt{|\Delta_q|^2+\tilde{\omega}_{n,\lambda}^{\pm2}}$, $\Phi^\pm_{\lambda,n}=i\omega_n L/(v_F\cos \theta)\mp\phi_\lambda/2$.

As there is no mixing between two bands, the total current of the junction is simply the sum of the currents from each of the two helical bands. From Eq. (\ref{current}) one can write
\begin{equation}
    \mathbf{I}_\text{tot}(\phi)=\sum_\lambda \left(1+\lambda\tilde{\alpha}\right) \mathbf{I}(\mathbf{q}_\lambda,\phi_\lambda). \label{Itot}
\end{equation}
where
\begin{equation}
    \mathbf{I}(\mathbf{q}_\lambda,\phi_\lambda)=  -\frac{i\pi W N_0 v_F T}{2}\sum_{\omega_n}\langle\text{Tr}[\tau_z g_\lambda\hat{\mathbf{k}}]\rangle_\text{FS}. \label{I2D}
\end{equation}
Substituting Eq. (\ref{Aeta}) to Eq. (\ref{I2D}), one obtains 
\begin{equation}
    I_x(\mathbf{q}_\lambda,\phi_\lambda)=-i\pi N_0 W v_F T\sum_{\omega_n}  \int_{-\pi/2}^{\pi/2} \frac{d\theta}{2\pi}(A_+^\lambda-A_-^\lambda)\cos\theta. \label{currentperDOS}
\end{equation}

The quasi-1D case where the normal metal has only a few conduction channels is of particular interest, for which Eq. (\ref{I2D}) becomes
\begin{equation}
    I^\text{1D}(\mathbf{q}_\lambda,\phi_\lambda)=-\frac{i\pi N_0 v_F T }{2 k_F}\sum_{\omega_n,j} \text{Tr}\left[\tau_z\left( g_{\lambda,j}^+-g_{\lambda,j}^- \right)\right].\label{I1D}
\end{equation}
Here we define $g_{\lambda,j}^{\pm}=g_\lambda(\mathbf{r},\hat{\mathbf{k}}^{\pm}_j,i\omega_n)$ where the transverse momentum satisifies quantization condition $\hat{\mathbf{y}}\cdot\hat{\mathbf{k}}^{\pm}_j=2\pi j/(Wk_F)$ ($j\in \mathbb{Z}$) due to confinement effect. In this work, we focus on the case where there is only one conduction channel. With Eq. (\ref{Aeta}) one obtains
\begin{equation}
    I^\text{1D}(\mathbf{q}_\lambda,\phi_\lambda)=-\frac{i\pi N_0 v_F T }{ k_F}\sum_{\omega_n}  [A_+^\lambda(\theta=0)-A_-^\lambda(\theta=0)].  \label{eq:I1D}
\end{equation}

\begin{figure}
    \centering
        \subfigure[]{    \centering
    \includegraphics[width=0.4\textwidth]{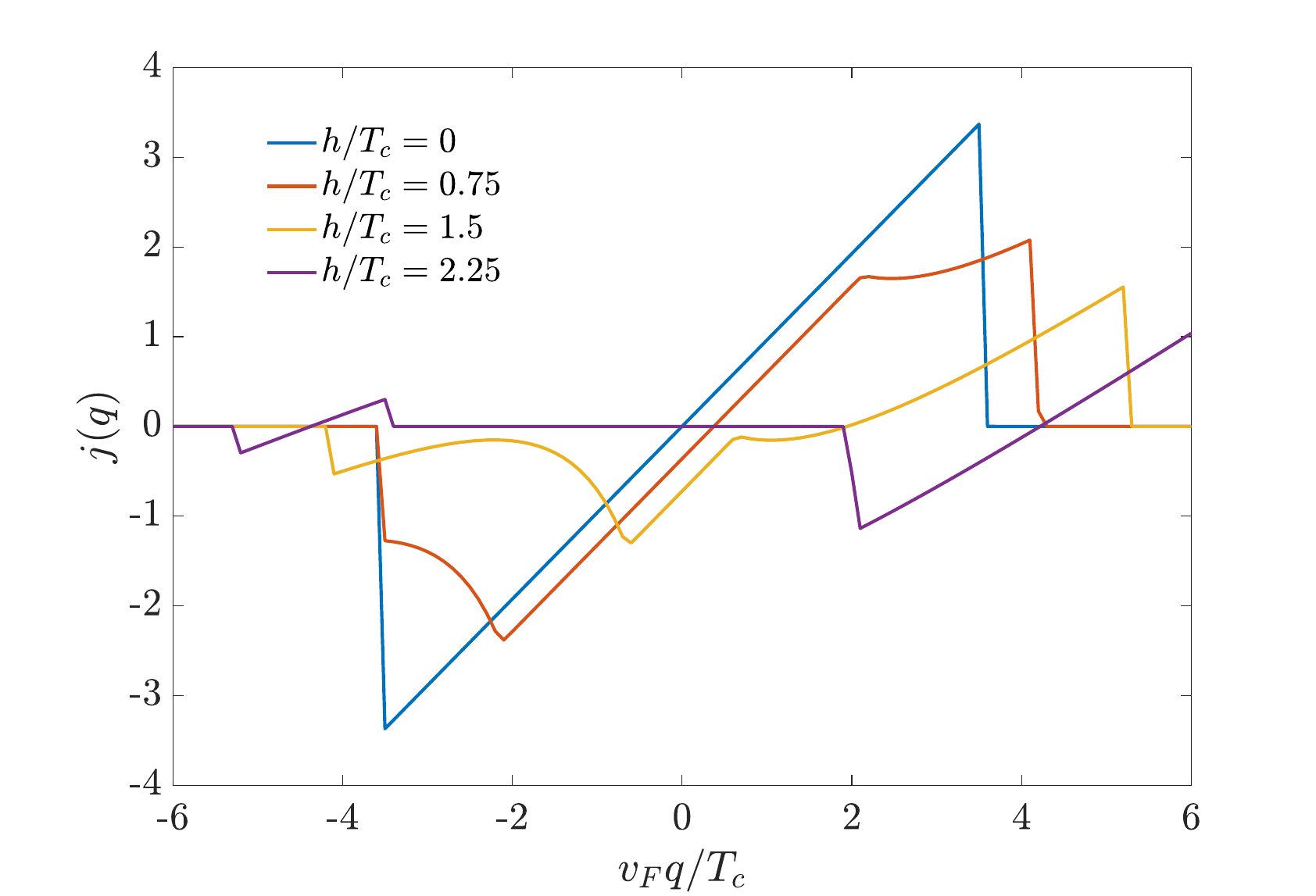}\label{fig:jvsq1}}
    \subfigure[]{    \centering
    \includegraphics[width=0.4\textwidth]{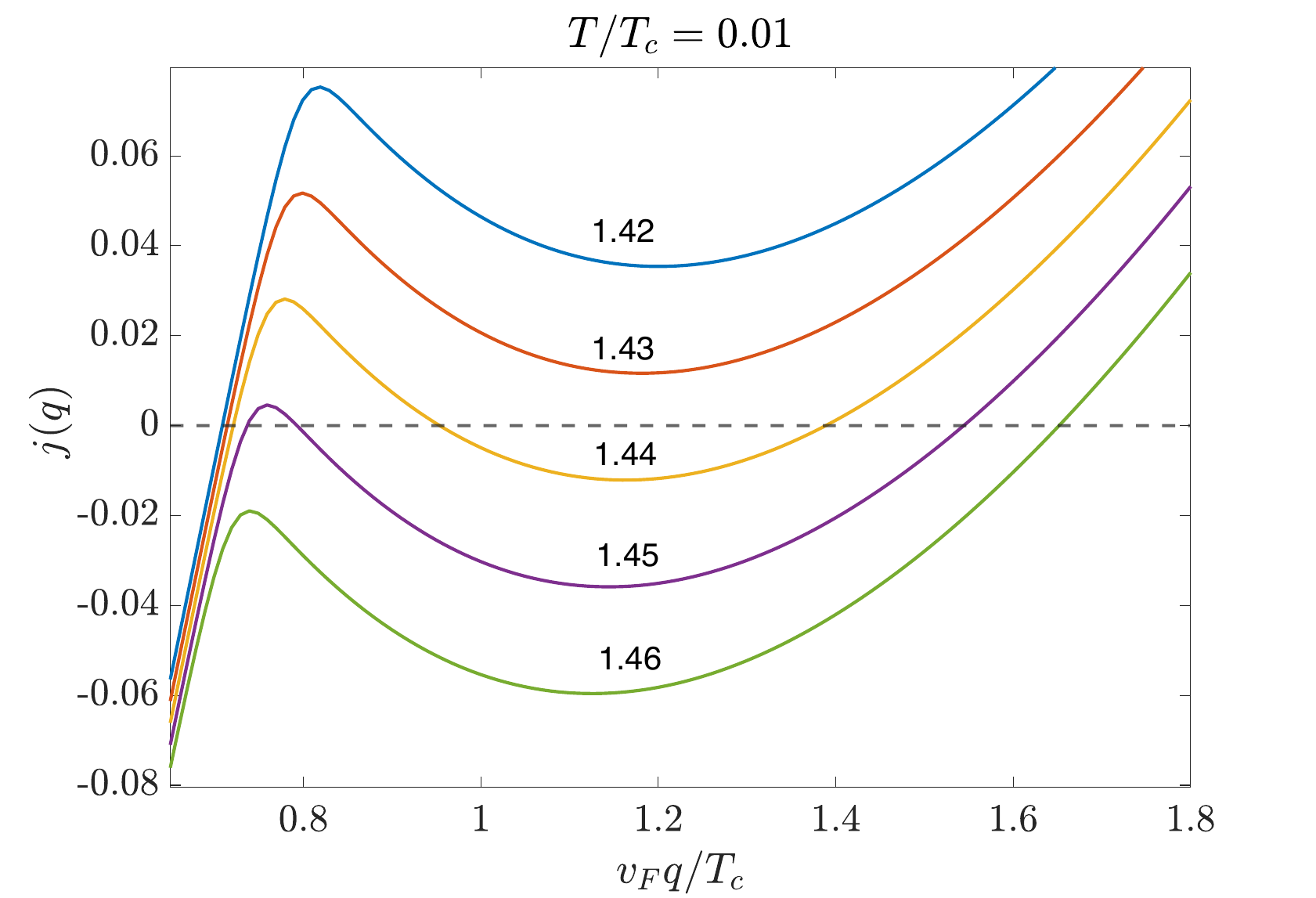}\label{fig:jvsq2}}
        \subfigure[]{    \centering
    \includegraphics[width=0.4\textwidth]{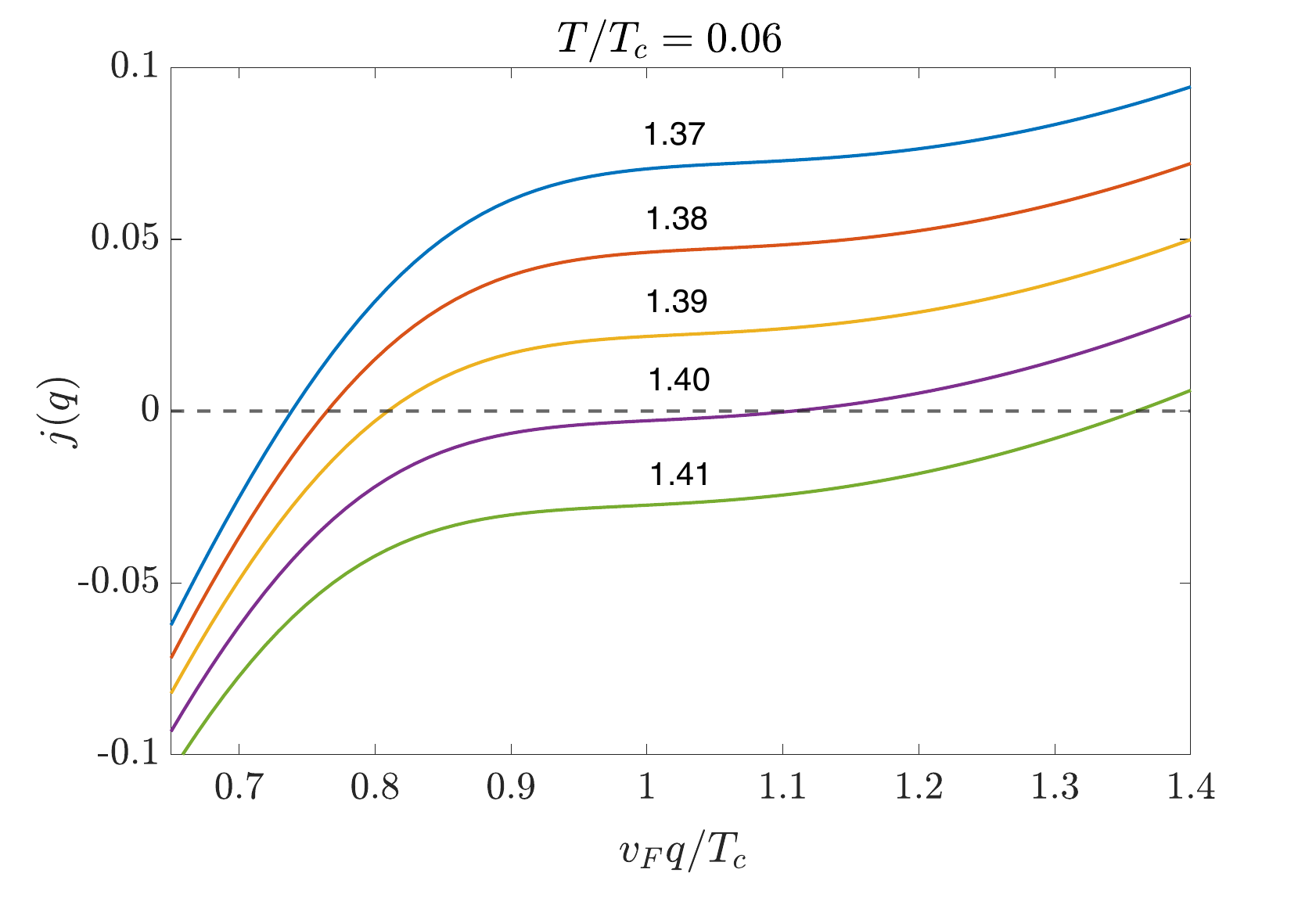}\label{fig:jvsq3}}
    \caption{(a) The supercurrent $j(q)$ vs Cooper pair momentum $q$ for different $h/T_c$ at $T/T_c=0.01$. (b-c) Zoomed-in plot of $j(q)$ near the first-order transition line between weak and strong helical phase at different temperatures, with different colored curves corresponding to different values of $h/T_c$ (indicated by the numbers in the plots).}
    \label{jvsq}
\end{figure}

\begin{figure}
\centering
        \subfigure[]{    \centering
    \includegraphics[width=0.4\textwidth]{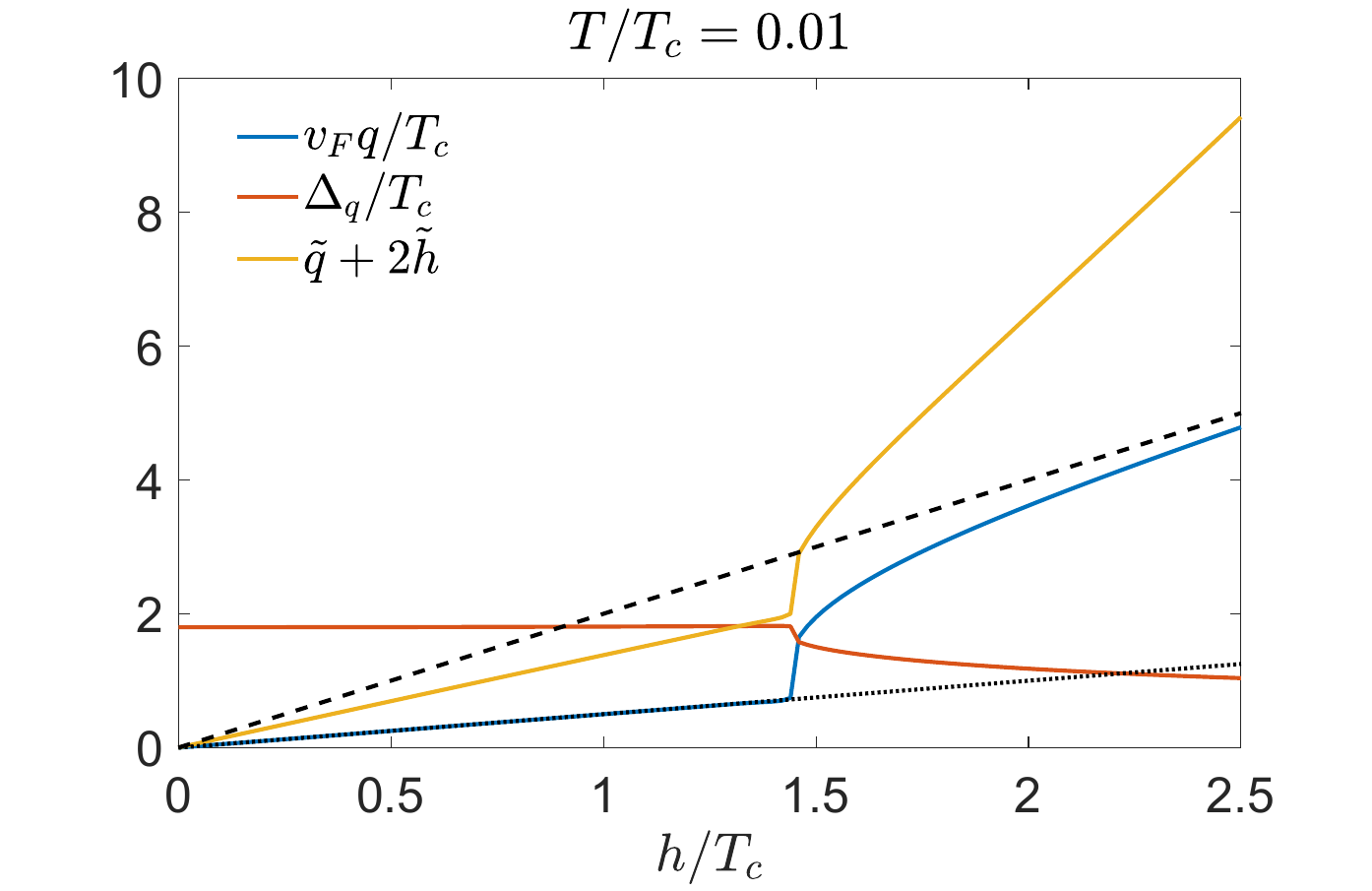}}
    \subfigure[]{    \centering
    \includegraphics[width=0.4\textwidth]{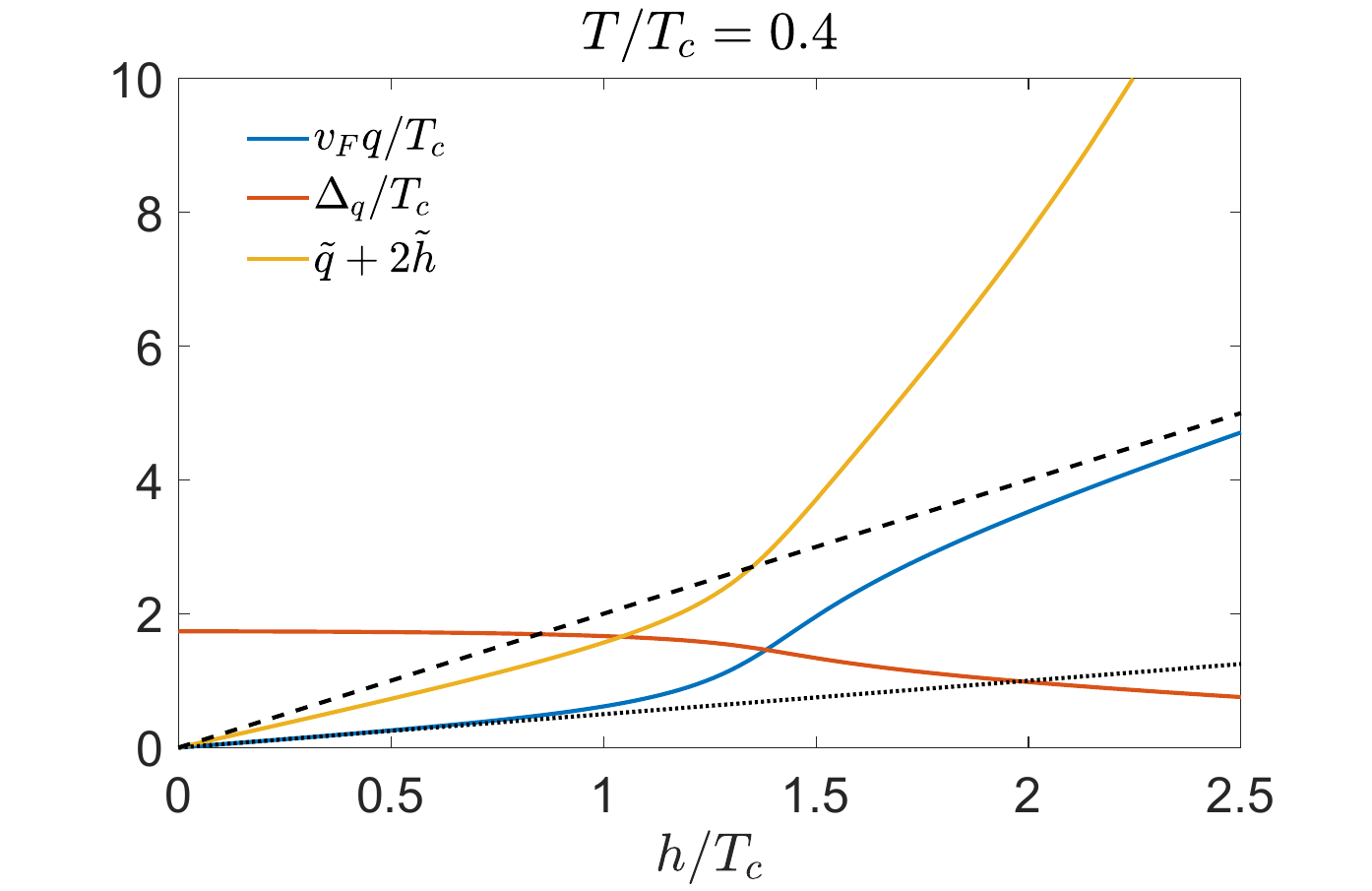}}
    \caption{The Cooper pair momentum $q$ and gap $\Delta_q$ as a function of magnetic field $h$ for a homogeneous bulk superconductor. The dashed and dotted line denotes the approximate expression $\tilde{q}=2\tilde{h}$ and $\tilde{q}=2\tilde{\alpha} \tilde{h}$ respectively.}
    \label{qvsh}
\end{figure}
\begin{figure}
    \centering
    \includegraphics[width=0.43\textwidth]{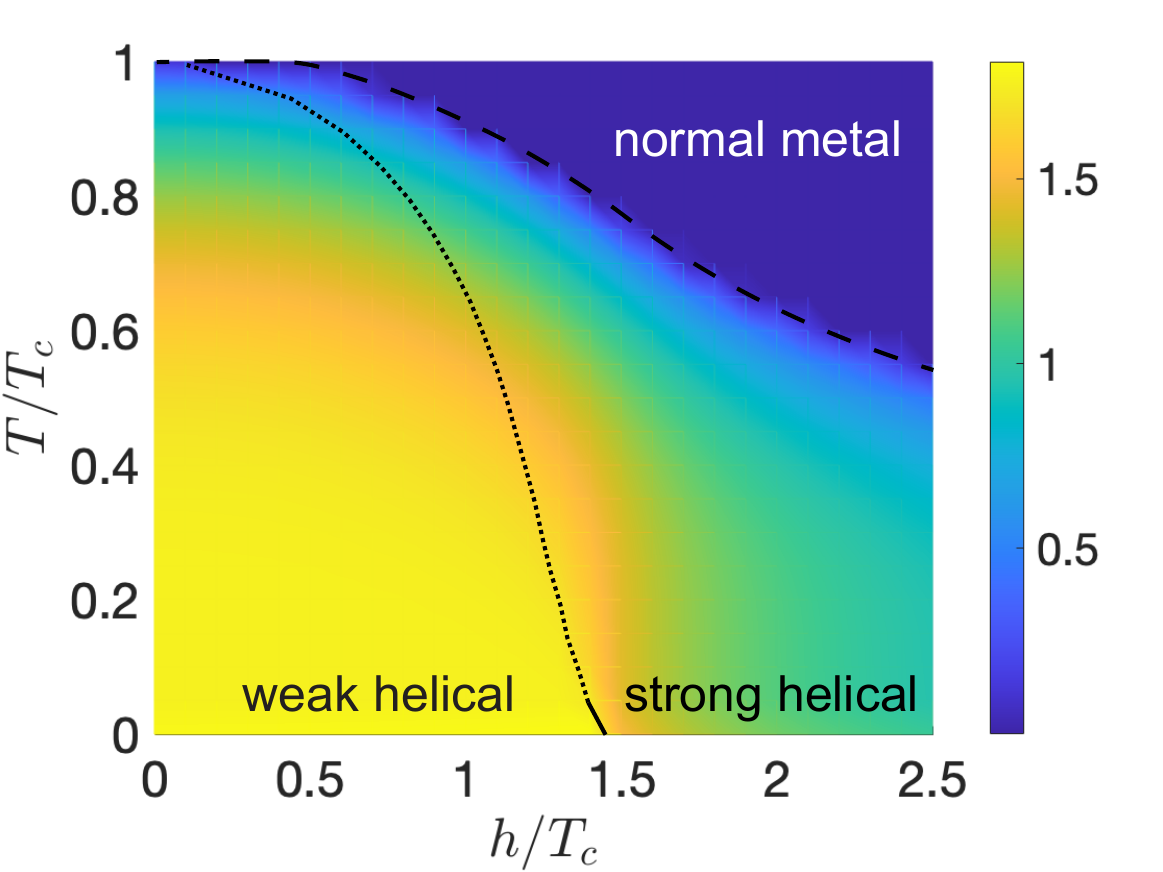}
    \caption{The phase diagram of bulk helical superconductor where the color denotes the magnitude of $\Delta_q/T_c$, for $\tilde{\alpha}=0.25$. The dashed line denotes the superconducting phase tranition, and the solid (dotted) line denotes the first-order (crossover) Lifshitz transition between the weak and strong helical phases. The crossover line is determined by $\tilde{q}+2\tilde{h}=2$ at which the superconducting gap of the $\lambda=-$ band closes.}
    \label{fig:phasediagram}
\end{figure}

\begin{figure*}[htp]
    \centering
        \subfigure[]{    \centering
    \includegraphics[width=0.3\textwidth]{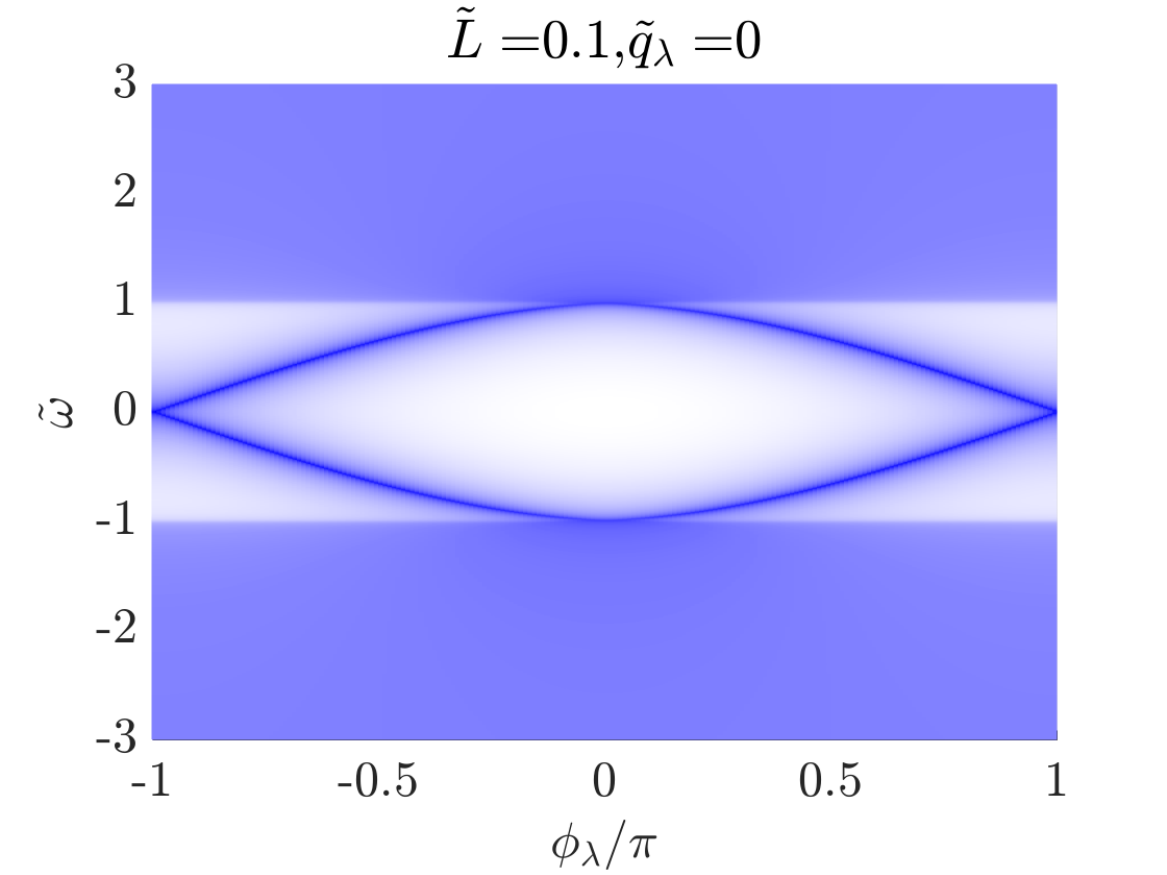}}
    \subfigure[]{    \centering
    \includegraphics[width=0.3\textwidth]{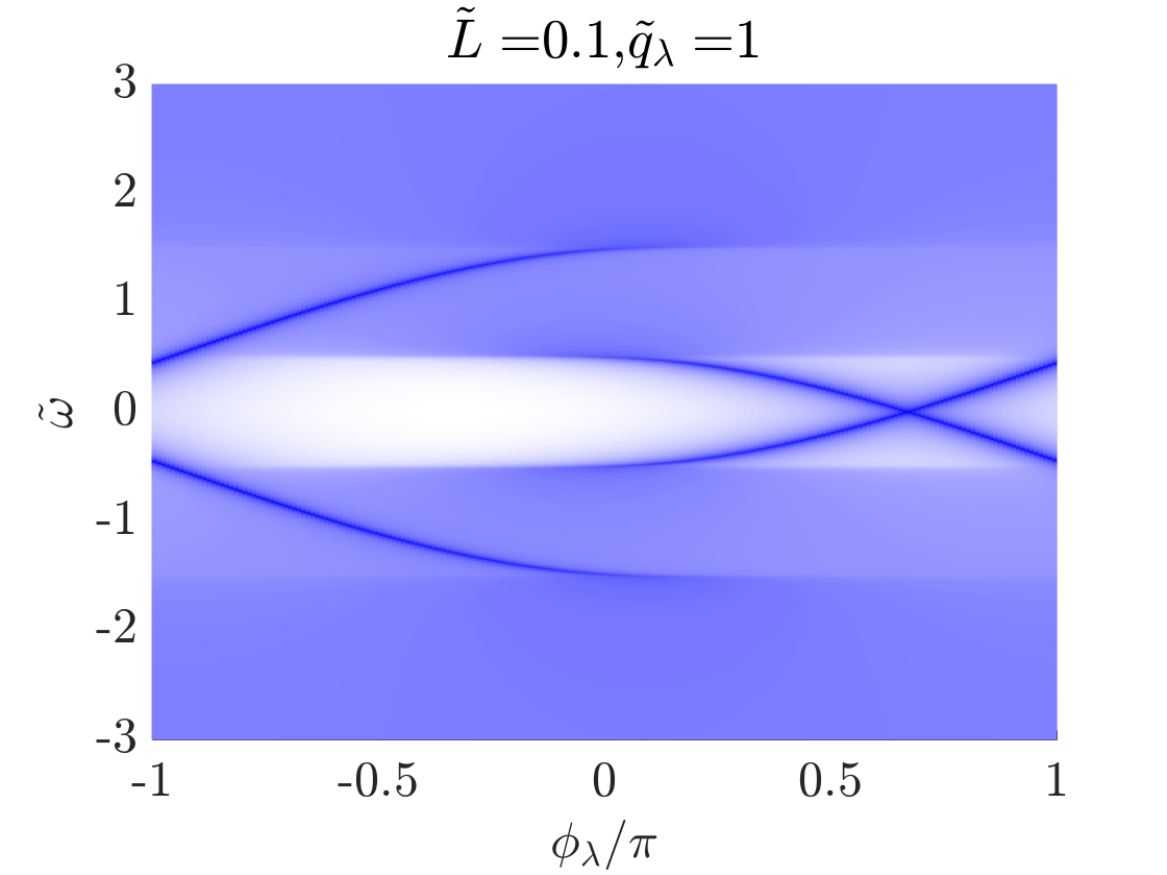}}
        \subfigure[]{    \centering
    \includegraphics[width=0.3\textwidth]{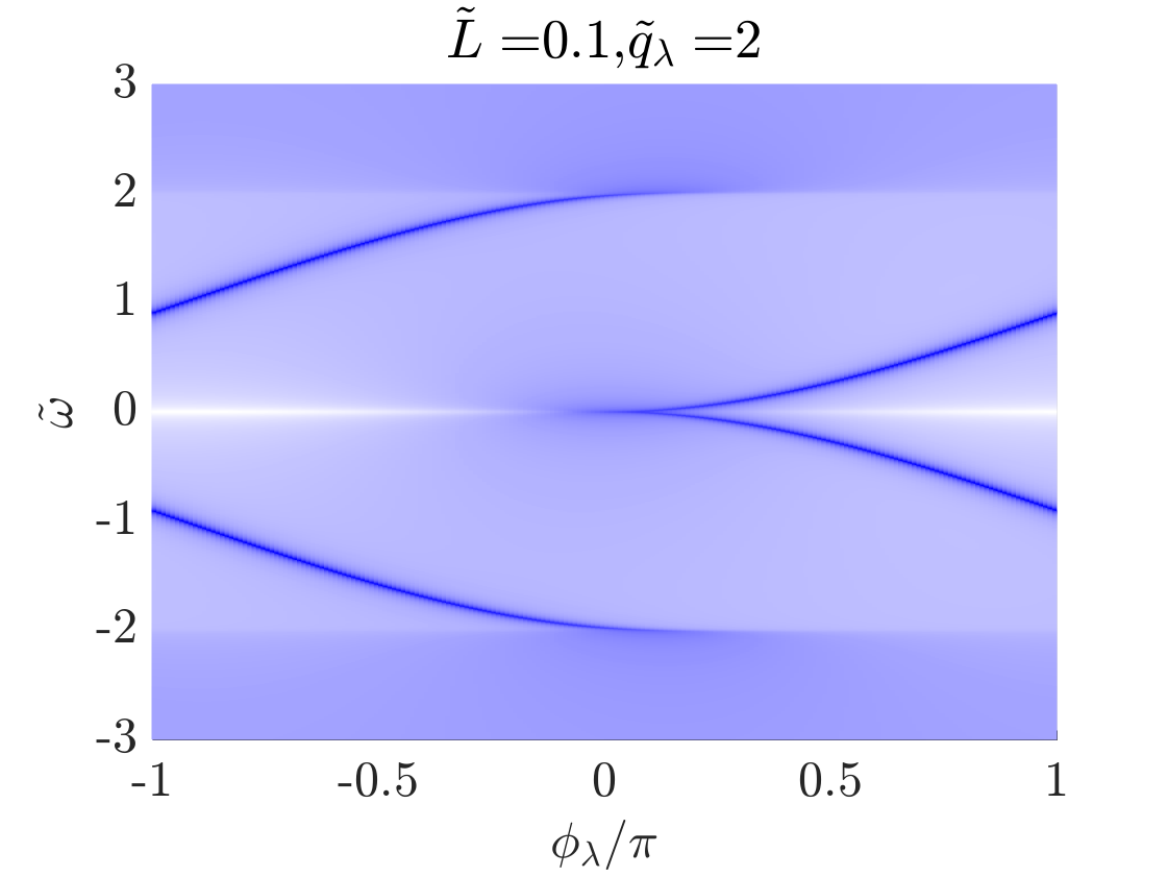}}
        \subfigure[]{    \centering
    \includegraphics[width=0.3\textwidth]{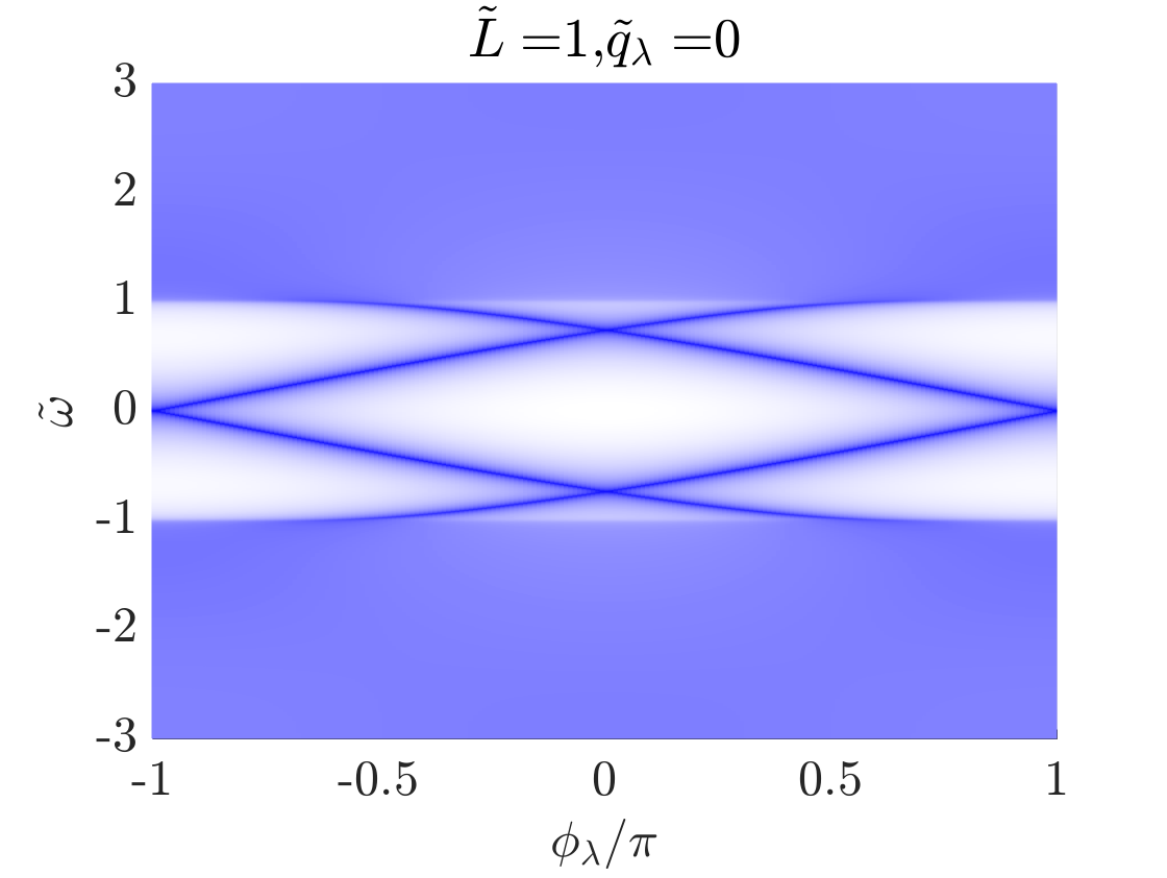}}
    \subfigure[]{    \centering
    \includegraphics[width=0.3\textwidth]{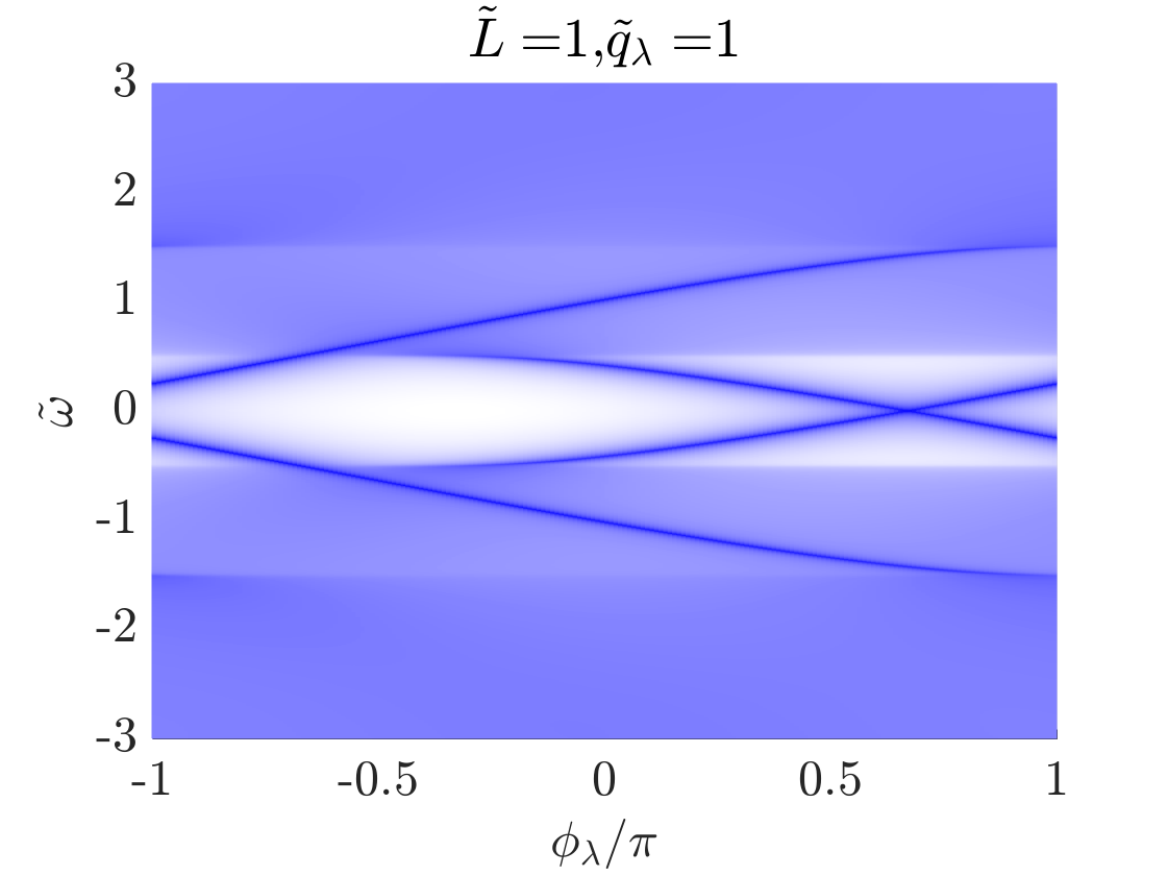}}
        \subfigure[]{    \centering
    \includegraphics[width=0.3\textwidth]{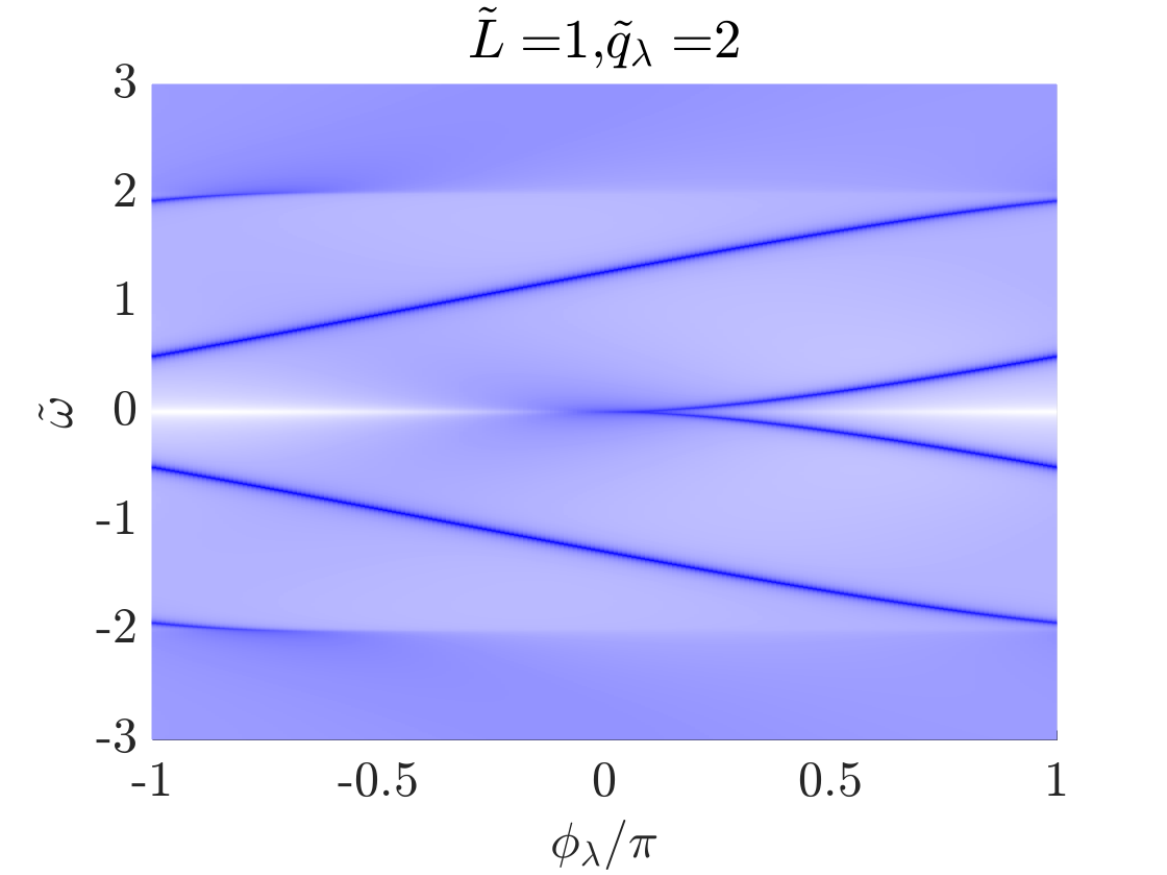}}
            \subfigure[]{    \centering
    \includegraphics[width=0.3\textwidth]{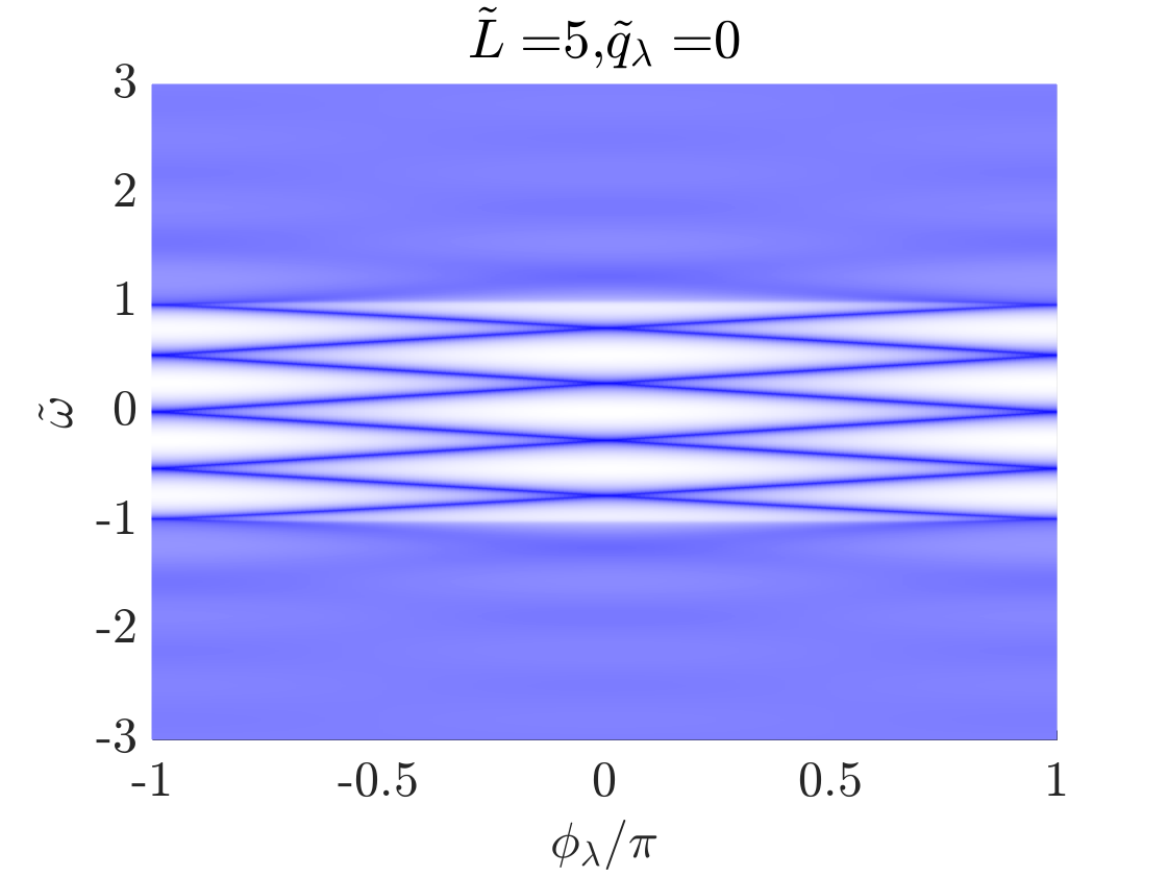}}
    \subfigure[]{    \centering
    \includegraphics[width=0.3\textwidth]{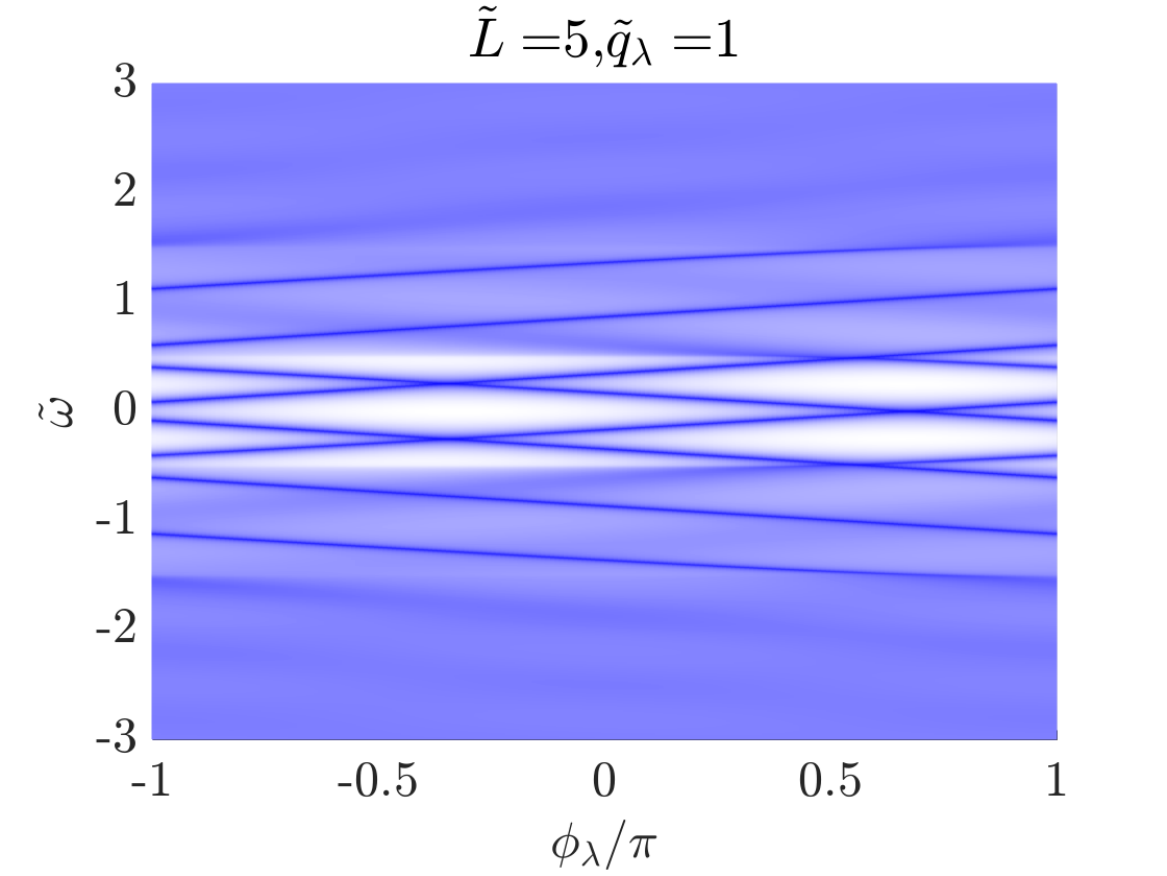}}
        \subfigure[]{    \centering
    \includegraphics[width=0.3\textwidth]{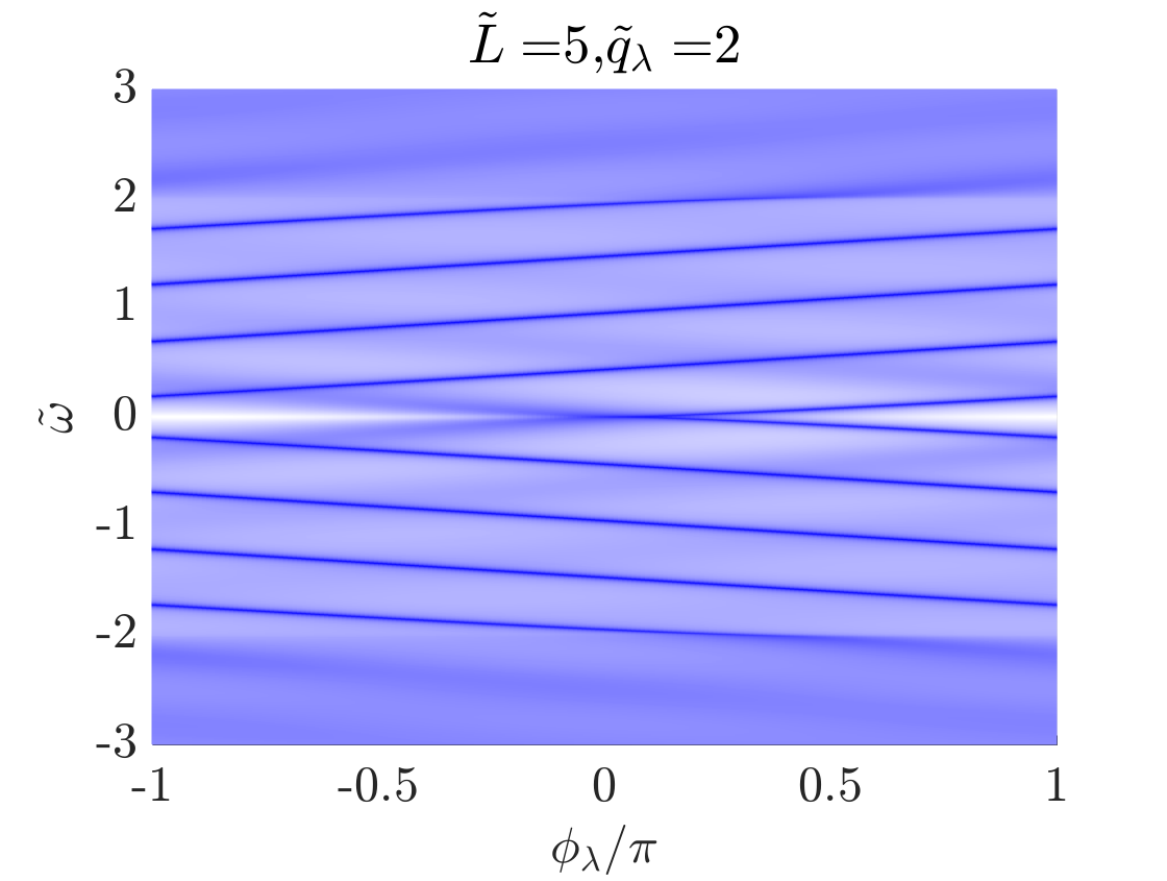}}
    
    \caption{The single-band energy spectrum in the normal region as a function of $\phi_\lambda$ for different $\tilde{L}$ and $\tilde{q}_\lambda$.
    The darker (lighter) color denotes higher (lower) local density of states.
    }
    \label{fig:LDOS}
\end{figure*}
\section{Results}\label{Sec:result}
In the following, we define dimensionless quantities $\tilde{\omega}=\omega/\Delta_q$, $\tilde{h}=h/\Delta_q$, $\tilde{q}=q\xi$, $\tilde{q}_\lambda=q_\lambda\xi$ and $\tilde{L}=L/\xi$, where $\xi=v_F/\Delta_q$ is the coherence length. Unless specified, we assume that $\mathbf{h}=h\hat{\mathbf{y}}$, $\mathbf{q}=q\hat{\mathbf{x}}$ and there is single channel in the normal metal region of SNS junction. We set $\tilde{\alpha}=0.25$ and we expect most of our results to hold qualitatively in general.

\subsection{Bulk SDE in the helical phase}\label{sec:bulk}

As demonstrated in the previous section, the Eilenberger equation can be solved self-consistently for the bulk case, and the bulk supercurrent $j(q)=\sum_\lambda j_\lambda$ can be obtained for different Cooper pair momentum $q$, as shown in Fig. \ref{jvsq} (see Ref. \cite{IlicBergeret22} and appendix therein for more details). As seen from the plot, SDE is present as long as $h/T_c\neq 0$.
The equilibrium state is determined by requiring $j(q)=0$,\footnote{There are two solutions at large $h$, and we choose the one with positive $q$ as it has lower free energy.} and the corresponding $q$ and $\Delta_q$ are hence obtained (plotted in Fig. \ref{qvsh}). We observe that both $q$ and $\Delta_q$ behave differently at low and high fields, between which there is a sudden jump or crossover. To understand these behaviors, we note that according to Eqs. (\ref{LDOS}) and (\ref{eq:homoSolution}) the angular-resolved LDOS is given by
\begin{equation}
    \nu^\text{ang}_\lambda=N_\lambda \text{Re}\left[\frac{-i (\tilde{\omega}-\tilde{q}_\lambda)\cos \theta/2)}{\sqrt{1-(\tilde{\omega}+i\delta-\tilde{q}_\lambda \cos \theta/2)^2}}\right],\label{eq:LDOS}
\end{equation}
which shows that the spectral gap range of band $\lambda$ is $|\tilde{\omega}|<1-|\tilde{q}_\lambda|/2=1-|(\tilde{q}-2\lambda \tilde{h})/2|$. When $h$ is weak $\tilde{q}_-\equiv|\tilde{q}+2\tilde{h}|<2$, the system is in a gapless phase, known as the weak helical phase, in which $\tilde{q}=2\tilde{\alpha} \tilde{h}$ \cite{DimitrovaFeigelman03}. When $\tilde{q}_-\equiv|\tilde{q}+2\tilde{h}|=2$, the system transitions to the strong helical phase in which $\tilde{q}\approx 2\tilde{h}$, accompanied by the vanishing superconducting gap of the $\lambda=-$ band and the emergence of the BFS \cite{DimitrovaFeigelman07, AgterbergKaur07}. 
The proliferation of Bogolyubov quasiparticles due to BFS results in the reduction of condensate density and hence the pairing amplitude $\Delta_q$, as seen in Fig. \ref{qvsh}, which eventually destroys the superconducting phase at large enough $h$.

The Lifshitz transition between the weak and strong helical phases is a first-order transition at low temperatures and the pairing momentum \(q\) jumps discontinuously, as shown in Fig. \ref{qvsh} (a) and observed in \cite{DaidoYanase22, Daido2022_2}, but becomes a crossover at higher temperatures as shown in Fig. \ref{qvsh} (b) and observed in \cite{Daido2022_2, IlicBergeret22}.
The change in the nature of the phase transition can be understood from the change in the form of $j(q)$ shown in Fig. \ref{jvsq}. At low temperatures, \(j(q)\) exhibits nonmonotonous behavior and has a local minimum, as shown in Fig. \ref{fig:jvsq2}. As the magnetic field increases, the local minimum moves downward and eventually crosses zero, indicating the formation of a new minimum of the free energy with \(j(q_s)=0\). This new minimum becomes a global minimum as the magnetic field is further increased, and the equilibrium value of \(q\) suddenly jumps from its value in the weak phase \(q_w\) to the new value of \(q_s\). As the temperature increases, the local minimum of \(j(q)\) becomes less pronounced and eventually becomes an inflection point at some temperature \(T_\text{CEP}\) which marks the critical end point (CEP) of the first-order phase transition, as shown in Fig. \ref{fig:jvsq3}. Above that temperature, \(q\) moves rapidly but continuously around the inflection point as a function of the magnetic field. 
In Fig. \ref{fig:phasediagram} we plot the phase diagram, and find that the critical end point at which the weak first-order transition line ends is located around $T_{\text{CEP}}/T_c\approx0.05$ for $\tilde{\alpha}=0.25$\footnote{We note that the first-order transition becomes continuous at larger $\tilde{\alpha}$, which results in the absence of CEP as well as the ideal diode efficiency.However it does not affect the results regarding the Josephson junctions.}.

Below \(T_\text{CEP}\), in Fig. \ref{fig:jvsq2} the lower critical current \(I_{c-}\) is generally determined by the local minimum of \(j(q)\) closest to \(q_s\) and not the global minimum of \(j(q)\): since lowering the current continuously below this value generally requires moving across a barrier in the free energy, transport likely becomes dissipative due to the formation of local domains are corresponding domain wall motion \cite{ChakrabortyBlackSchaffer24}. \(|I_{c-}|\) is smallest close to the first-order transition, and as the CEP is approached along this transition line \(|I_{c-}|\) eventually approaches zero when the minimum vanishes. Formally, then, the diode coefficient \(\eta=(I_{c+}+I_{c-})/(I_{c+}-I_{c-})\) approached its nominally `perfect' value of unity. Similar perfect diode efficiency has been found to occur near the tricritical point of the FFLO state \cite{YuanFu22} and second-order phase transitions between uniform and non-uniform SCs \cite{ShafferChichinadzeLevchenko24, ChakrabortyBlackSchaffer24}. We thus identify another type of critical point that also gives rise to a potentially perfect SDE. However, we note that at the CEP the barrier in the free energy, which gives rise to the dissipation and the resulting perfect diodicity, also vanishes, and as a result precisely at the CEP the lower critical current \(J_{c-}\) should instead be determined by the next nearest local minimum of \(j(q)\) (which coincides with the global minimum in our case). In practice, therefore, we expect that the perfect value of the diode coefficient is never reached but rather reaches some maximum value once the free energy barrier is sufficiently low that tunneling can take place to the new minimum without dissipation. The value of this minimal barrier is determined by details of the system that are beyond our simplified model, but we nevertheless expect that in principle the diode efficiency should be strongly enhanced around the CEP.

\subsection{SNS junction}
\subsubsection{Single-band properties}\label{Sec:singleband}

\begin{figure*}[ht]
\centering
    \subfigure[]{ \centering
    \includegraphics[width=0.4\textwidth]{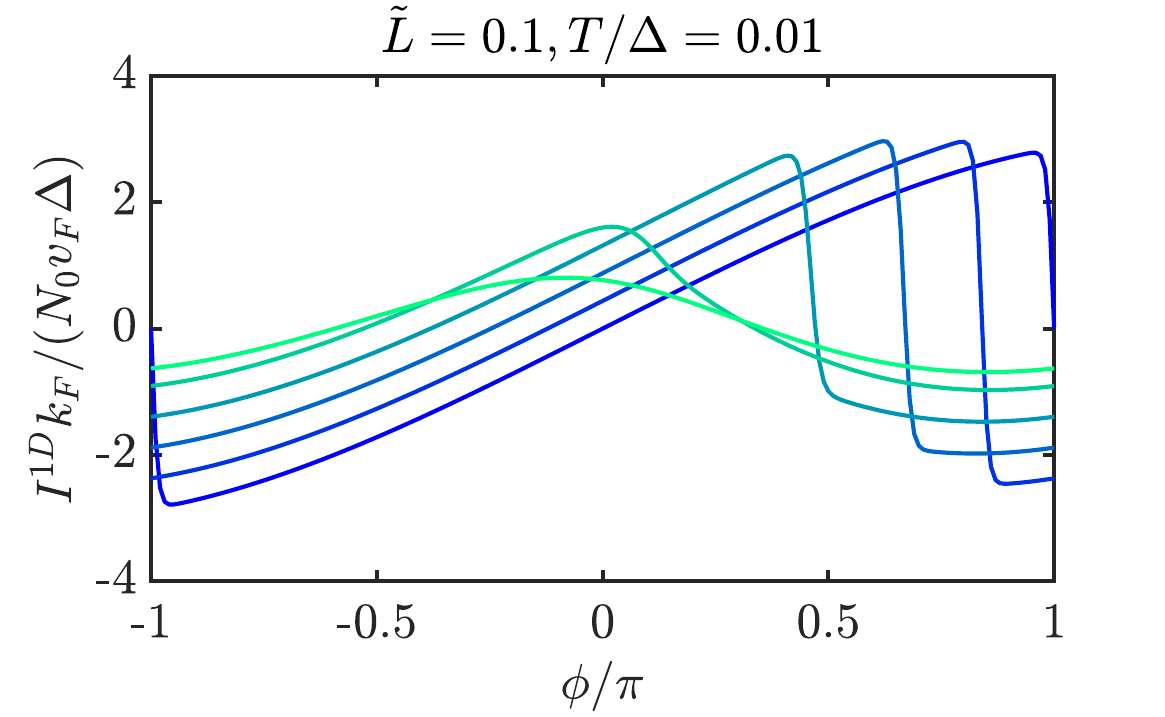}\label{fig:I1D1}}
\subfigure[]{    \centering
    \includegraphics[width=0.4\textwidth]{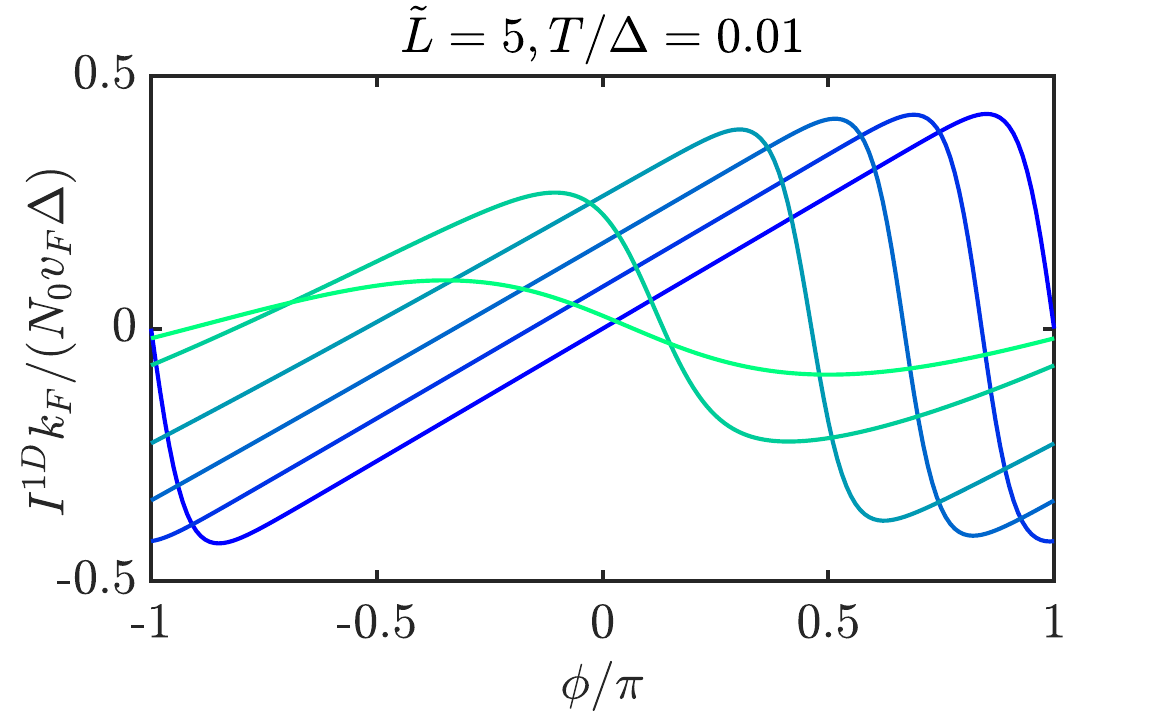}\label{fig:I1D2}}
    \subfigure[]{    \centering
    \includegraphics[width=0.4\textwidth]{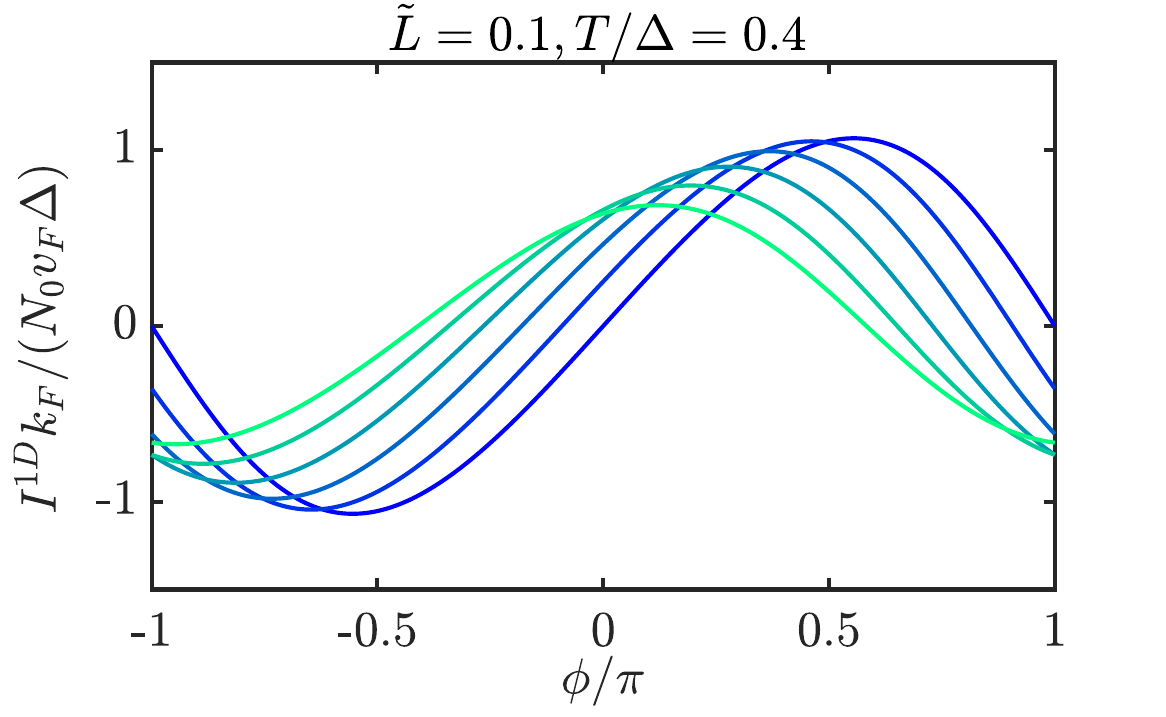}\label{fig:I1D3}}
        \subfigure[]{    \centering
    \includegraphics[width=0.4\textwidth]{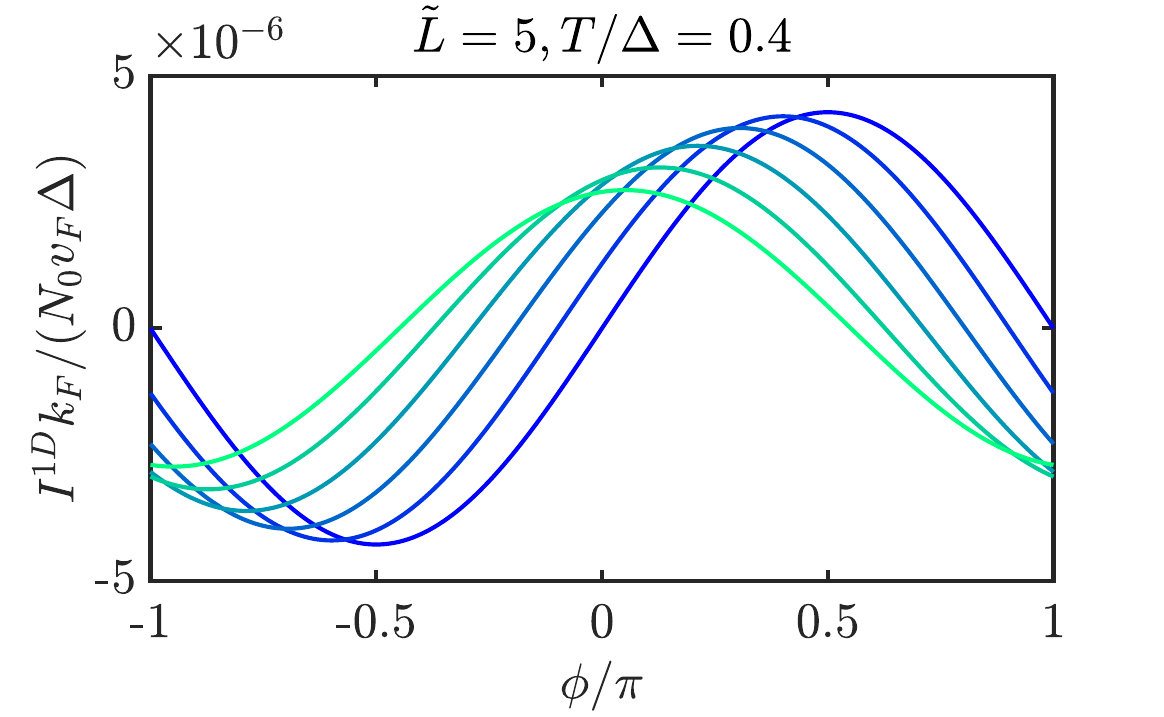}\label{fig:I1D4}}
    \caption{The single-band CPR $I^\text{1D}(q_\lambda,\phi_\lambda)$ for different $\tilde{L}$, $T/\Delta$ and $\tilde{q}_\lambda$. The color from blue to green correspond to $\tilde{q}_\lambda=0,0.5,1,1.5,2,2.5$. }
    \label{fig:I1D}
\end{figure*}

In an SNS junction, discrete levels of Andreev bound state are formed in the normal metal due to the Andreev reflection \cite{Andreev1964}. In the strong spin-orbit coupling limit, as the two helical bands decouple, the Andreev levels of each band do not mix and can be treated separately. For a single band $\lambda$, its structure as a function of $\tilde{\omega}$, $\tilde{q}_\lambda$, $\phi_\lambda$ and $\tilde{L}$ is shown in Fig. \ref{fig:LDOS}. The energy spectrum of the Andreev bound states along the $\pm$ direction is given by the poles of Eq. (\ref{Aeta}), namely
\begin{equation}
    \tilde{\omega}\tilde{L}\mp \phi_\lambda/2=\arccos{(\tilde{\omega}\mp \frac{\tilde{q}_\lambda}{2})}+n\pi,
\end{equation}
where $n\in \mathbb{Z}$.
For a short junction $\tilde{L}\ll 1$, the energy-phase relation of Andreev level is approximately given by
\begin{equation}
    \tilde{\omega}_+=\frac{\tilde{q}_\lambda}{2}+\cos\frac{\phi_\lambda}{2}, \quad\phi_\lambda\in[-2\pi,0),
\end{equation}
\begin{equation}
    \tilde{\omega}_-=-\frac{\tilde{q}_\lambda}{2}+\cos\frac{\phi_\lambda}{2}, \quad\phi_\lambda\in[0,2\pi).
\end{equation}
In the long junction limit $\tilde{L}\gg 1$, the low-lying Andreev states near the gap center $|\tilde{\omega}\mp \frac{\tilde{q}_\lambda}{2}|\ll 1$ are well described by 
\begin{equation}
    \tilde{\omega}_\pm\approx \frac{(2n+1)\pi\pm(\phi_\lambda+\tilde{q}_\lambda)}{2(\tilde{L}+1)}, \label{E-phase}
\end{equation}
from which one can see that the effect of non-vanishing $\tilde{q}_\lambda$ is to introduce a phase shift of the Andreev levels. 

The single-band current-phase relation (CPR) $I^\text{1D}(q_\lambda,\phi_\lambda)$ can be calculated using Eqs. (\ref{Aeta}) and (\ref{eq:I1D}). At low temperature, our result shown in Fig. \ref{fig:I1D1} for the short junction agrees well with that obtained using the scattering matrix formalism \cite{DavydovaFu22}. For a long junction at low temperature, we observe from Fig. \ref{fig:I1D2} that the single-band CPR can be roughly approximated by $I^\text{1D}(q_\lambda,\phi_\lambda)\approx I^\text{1D}(0,\phi_\lambda+\tilde{q}_\lambda)$ for small $q_\lambda$. This suggests that while there is an AJE that can be seen in the single-band CPR, there is only a weak JDE.  This can be understood by noting that $I^\text{1D}=2e(dE_\text{tot}/d\phi)$, where the total energy of the junction $E_\text{tot}$ is contributed by both the continuum and Andreev bound state. As seen from Eq. (\ref{E-phase}), the energy of the low-lying Andreev states is a function of $\phi_\lambda+\tilde{q}_\lambda$, the CPR is thus a function of $\phi_\lambda+\tilde{q}_\lambda$ as well, given the energy contribution of the high-lying Andreev levels and the continuum is insensitive to phase when $\tilde{q}_\lambda\ll 1$. When $\tilde{q}_\lambda\gtrsim 2$, the superconducting gap closes as a BFS forms, leading to significant suppression of Josephson current and asymmetric behavior of $I^\text{1D}$. At high temperatures, as indicated by Figs. (\ref{fig:I1D3}) and (\ref{fig:I1D4}), the single-band CPR becomes close to the standard trigonometric form. Moreover, when the temperature $T$ becomes comparable to the Thouless energy $E_T=v_F/L$, the Josephson current for a long junction is significantly suppressed (Fig. \ref{fig:I1D4}).
\begin{figure}
    \centering
    \includegraphics[width=0.45\textwidth]{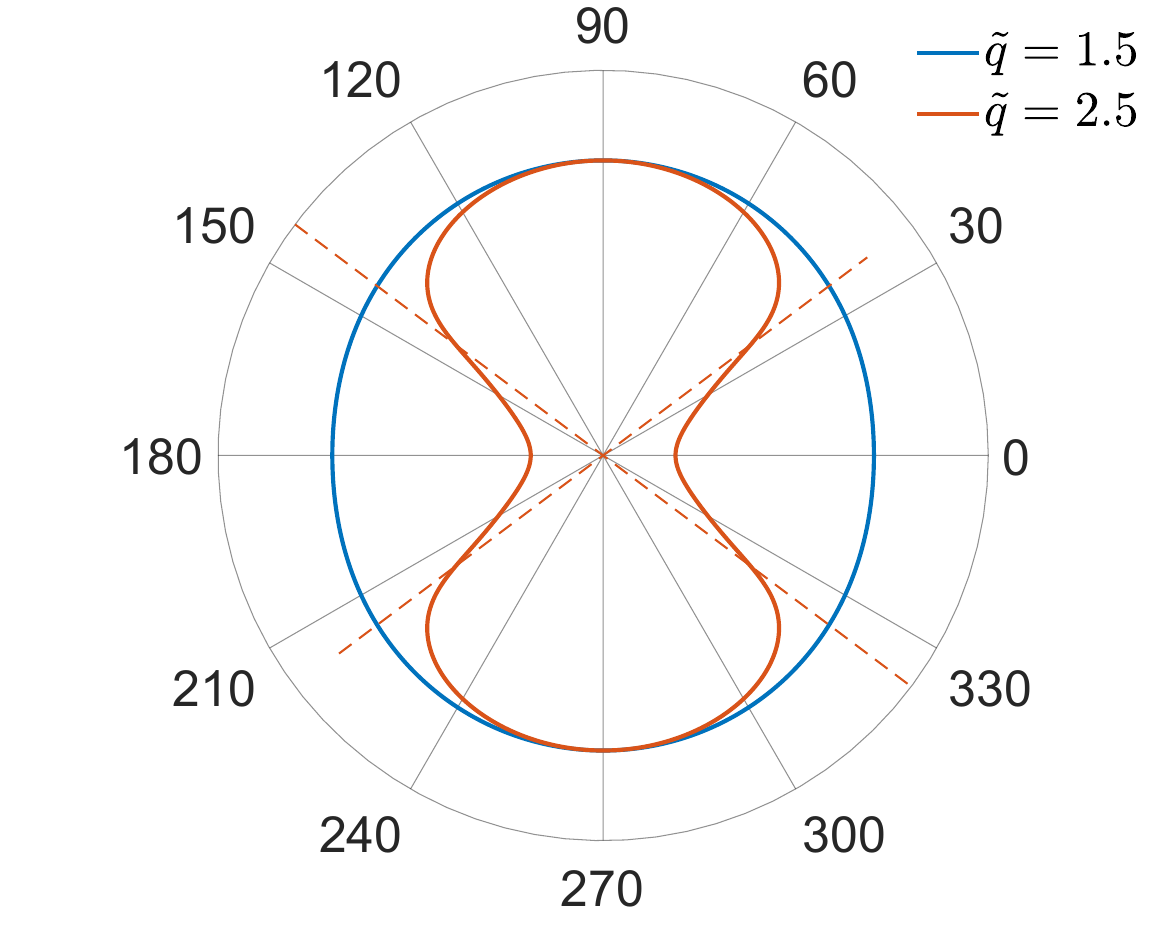}
    \caption{The angular dependence of the root-mean-square current $\sqrt{\overline{(I^{1D})^2}}/\Delta$ for $\tilde{q}=1.5$ (weak helical phase) and $\tilde{q}=2.5$ (strong helical phase) at $T/\Delta=0.01$. The red dashed line denotes the angle at which the BFS appears along the $x$-direction for $\tilde{q}=2.5$.}
    \label{fig:Jpolar}
\end{figure}

So far we have only considered the case where $\mathbf{q}_\lambda$ is along the $x$-direction. For the general case where there is an angle $\theta_h$ between $\mathbf{q}_\lambda$ and $x$-axis, the single-band CPR has the same form with substitution $q_\lambda\rightarrow q_\lambda\cos \theta_h$, as seen from Eq. (\ref{Aeta}). Physically, this is because the particles in the quasi-1D normal metal can only experience the superconducting gap along the $x$-direction, which directly depends on $q_\lambda\cos \theta_h$ (see Eq. (\ref{eq:LDOS})). In the strong helical phase, there is therefore a range of angle in which the superconductor is gapped along the $x$-direction and hence the suppression of Josephson current is almost absent. This leads to the strong anisotropy of Josephson current, shown in Fig. \ref{fig:Jpolar}.

\begin{figure*}[ht]
\centering
    \subfigure[]{ \centering
    \includegraphics[width=0.32\textwidth]{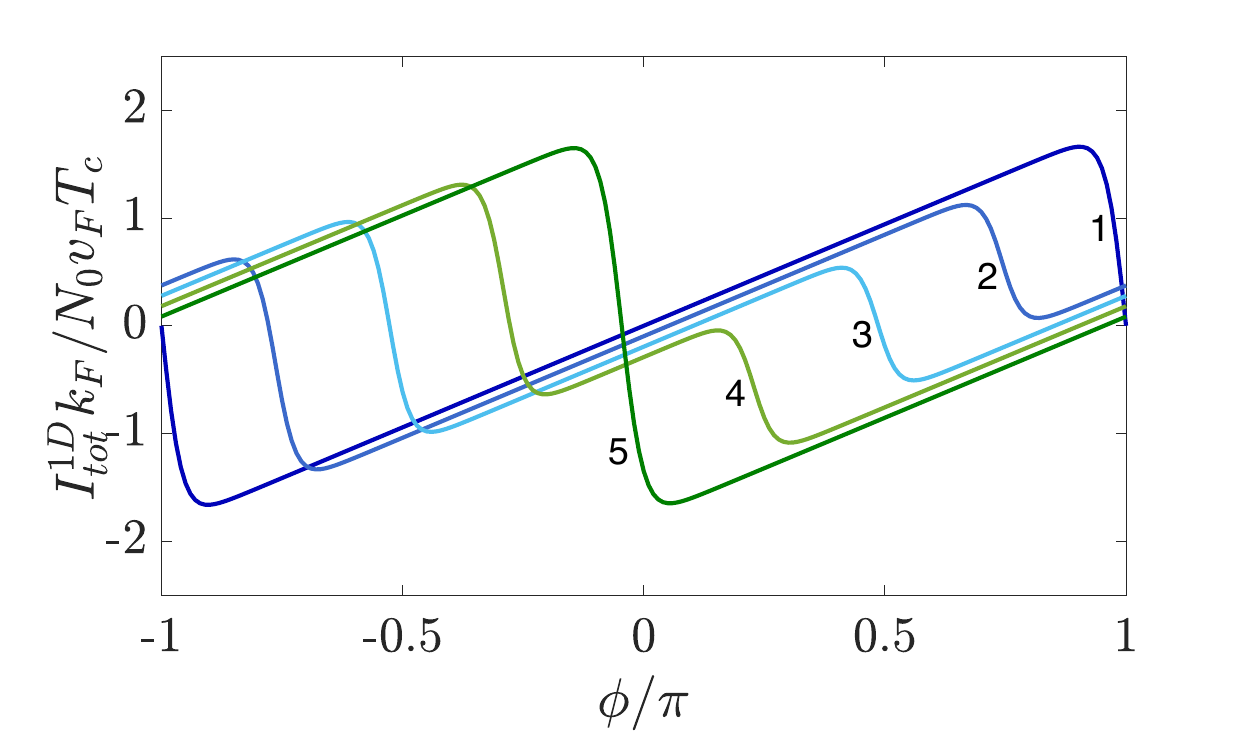}\label{fig:CPR1}}
\subfigure[]{    \centering
    \includegraphics[width=0.32\textwidth]{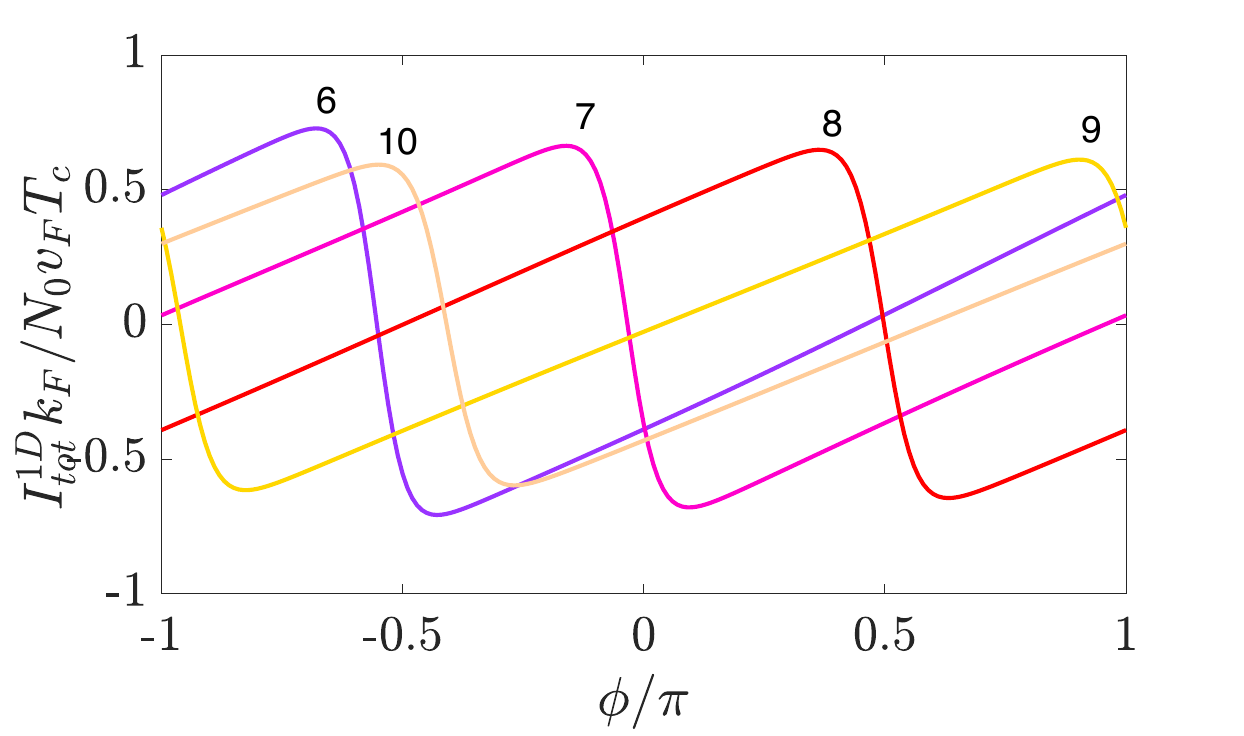}\label{fig:CPR2}}
        \subfigure[]{    \centering
    \includegraphics[width=0.32\textwidth]{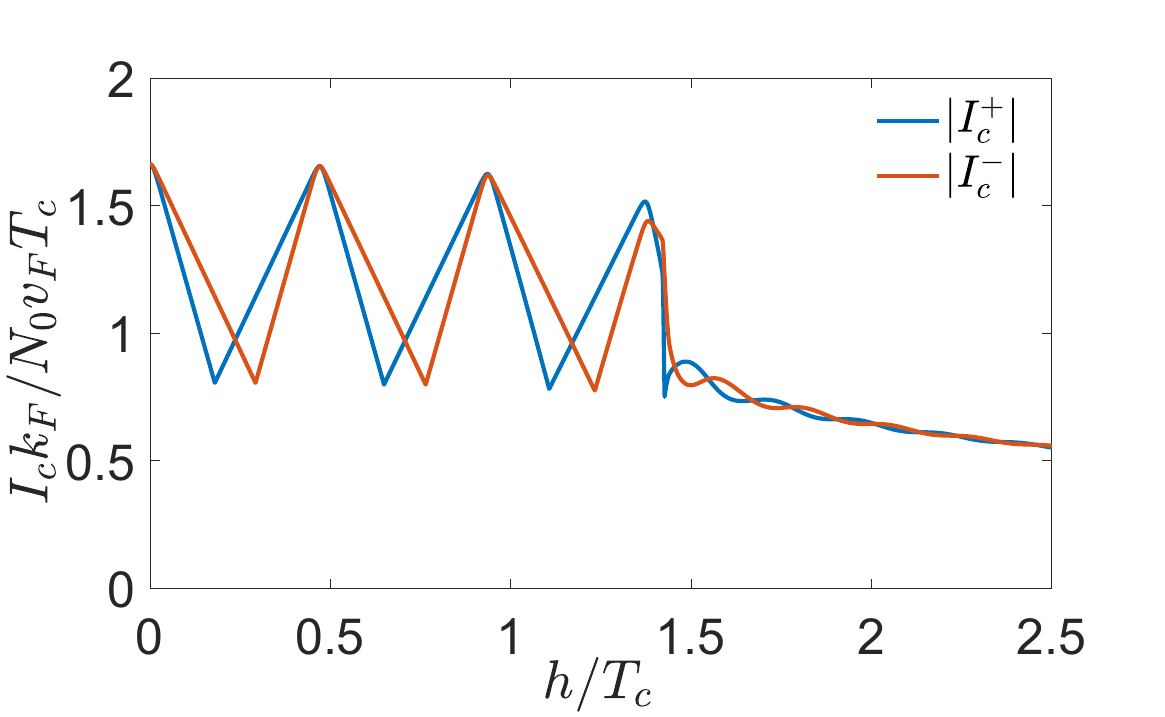}\label{fig:Ic}}
    \subfigure[]{    \centering
    \includegraphics[width=0.32\textwidth]{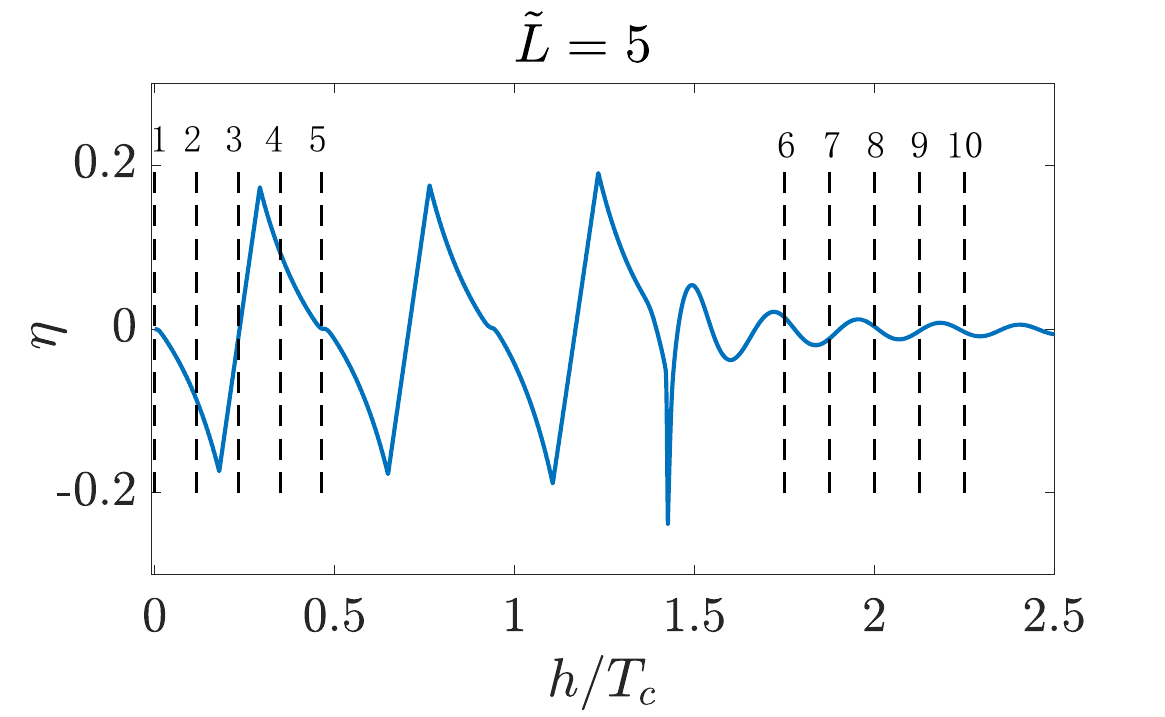}\label{fig:eta}}
    \subfigure[]{    \centering
    \includegraphics[width=0.32\textwidth]{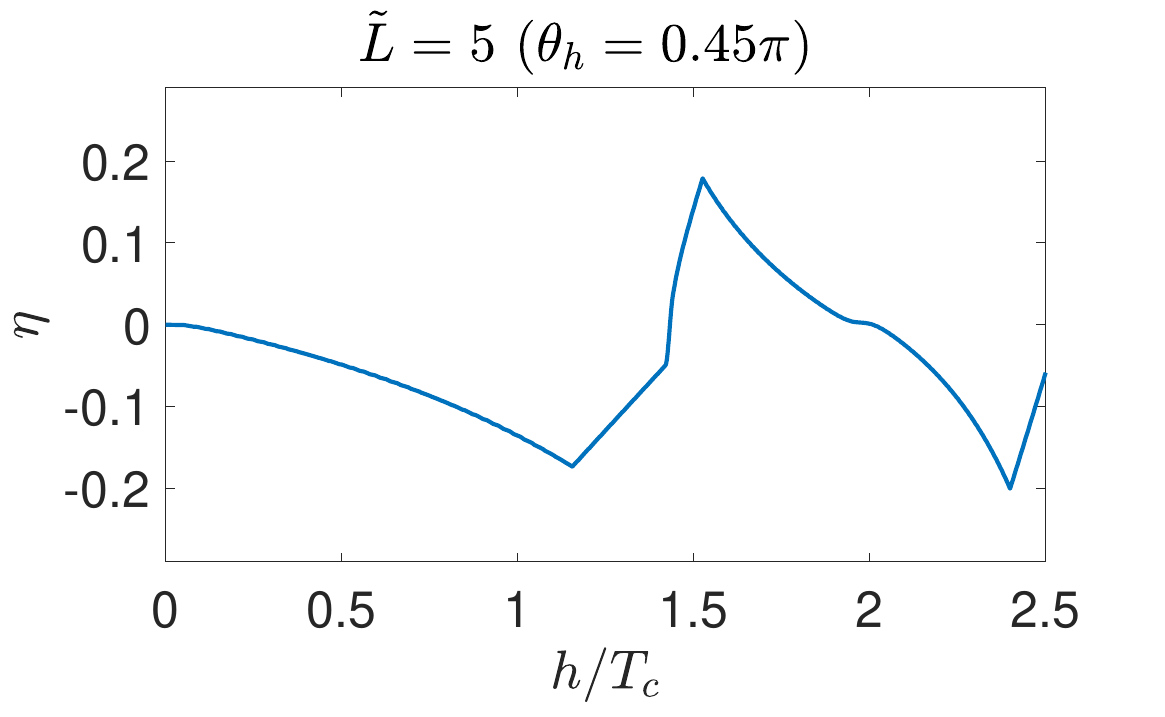}\label{fig:eta3}}
        \subfigure[]{    \centering
    \includegraphics[width=0.32\textwidth]{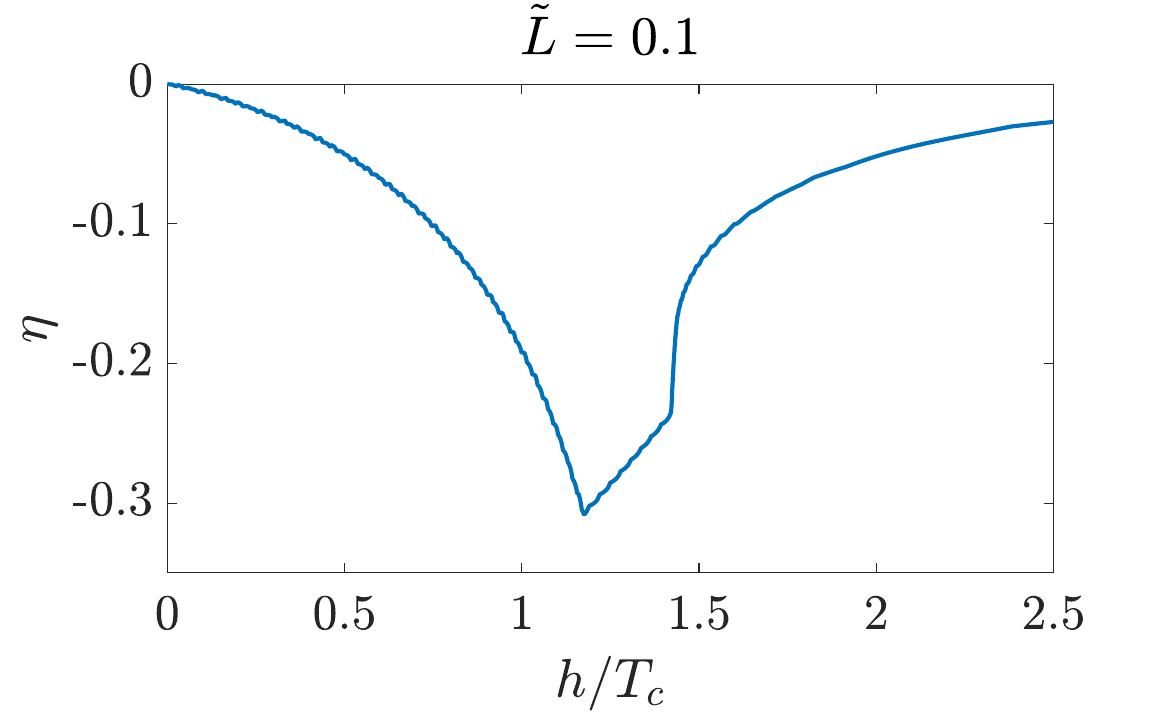}\label{fig:eta2}}

    \caption{Long junction $\tilde{L}=5$: (a)(b) The total current-phase relation for selected in-plane field in (d); (c)The critical current $I_c^{\pm}$ as a function of magnetic field $h$; (d)(e) The diode efficiency $\eta$ versus $h$ for different magnetic field orientations. Short junction $\tilde{L}=0.1$: (f) The diode efficiency $\eta$ versus $h$. The temperatures in all plots are set to $T/T_c=0.01$ and the magnetic field direction is $\theta_h=0$ except for (e).}
    \label{fig:Icetavsh}
\end{figure*}

\subsubsection{Multi-band properties and JDE}

From the symmetry point of view, both time-reversal (TR) and inversion symmetry breaking are necessary for the existence of JDE. In the setup considered in this work, the presence of Rashba SOC breaks the inversion symmetry and leads to the splitting of the helical bands, which is reflected in the different band DOS $N_\lambda=N_0(1+\lambda\tilde{\alpha})$. Without the in-plane field, the system is TR-invariant and shows no JDE, which can be seen by noting that the TR operation transforms $I^\text{1D}(q_\lambda,\phi_\lambda)\rightarrow I^\text{1D}(-q_\lambda,-\phi_\lambda)=-I^\text{1D}(q_\lambda,\phi_\lambda)$. However, when the magnetic field is present, the TR operation transforms $q_\lambda\rightarrow -q_\lambda-4Q_\lambda$ and $\phi_\lambda\rightarrow -\phi_\lambda-4Q_{\lambda x}L$, making the JDE possible. Nevertheless, it is not guaranteed that TR and inversion symmetry breaking would always give rise to JDE. Recall from Eq. (\ref{Itot}) that the total CPR is the sum of that from each bands weighted by density of states, which can be expanded as series of harmonics
\begin{equation}
    I^{1D}_\text{tot}(\phi)=\sum_\lambda(1+\lambda \tilde{\alpha})\sum_{n=1}^\infty D^\lambda_n\sin(n\phi+\gamma^\lambda_n).
\end{equation}
If only the lowest harmonic $n=1$ is present, only AJE, i.e. $I^{1D}_\text{tot}(\phi=0)\neq0$, and no JDE is present, even when both symmetries are broken. Therefore, non-vanishing higher harmonics in single-band CPR, which naturally arise in a ballistic junction at low temperatures (see Fig. \ref{fig:I1D}) \cite{Golubov2004}, are crucial for the existence of JDE \cite{GoldobinBuzdin07, YokoyamaNazarov14, IlicBergeret24, MeyerHouzet24}, as shown below.

In a long junction, as discussed earlier, the total current at small field can be approximated as
\begin{align}\label{I1Dapprox}
I_\text{tot}^{1D}&=\sum_\lambda\left(1+\lambda\tilde{\alpha}\right) I^\text{1D}(q_\lambda,\phi_\lambda)\\ \nonumber
    &\approx\sum_\lambda \left(1+\lambda\tilde{\alpha}\right) I^\text{1D}(0,\phi_\lambda+\tilde{q}_\lambda). 
\end{align}
The main effect of a weak magnetic field in a long junction is thus to introduce a relative phase shift between two bands
\begin{align}\label{eq:deltaphi}
\delta \phi&=\delta\phi^n+\delta\phi ^s\nonumber\\
&=(\phi_+-\phi_-)+(\tilde{q}_+-\tilde{q}_-)
=-4\tilde{h}(\tilde{L}+1).    
\end{align}
The first part $\delta \phi^n$ originates from the magnetic field in the normal metal region, which shifts the Fermi surface of different bands in the opposite direction. This leads to an accumulated relative phase shift when the particle and hole are reflected back-and-forth. The nonzero Cooper pair momentum $\tilde{q}_\lambda$ in the superconductor is responsible for the second phase shift $\delta \phi^s$, which is approximate and negligible compared to $\delta\phi^n$ in a long junction. Due to this relative shift, the upper and lower critical currents $I_c^\pm$, and consequently the diode efficiency $\eta$ defined below, oscillate as a function of applied magnetic field \(h\) and the diode efficiency switches sign multiple times. The period can be roughly obtained with Eq. (\ref{eq:deltaphi}), which approximately equal to \(\pi RT_c/2(\tilde{L}+1)\), where $R=\Delta_q/T_c\approx1.8$ is the gap ratio in the weak helical phase, close to $R_\text{BCS}\approx1.76$ for a BCS superconductor.  
In the limit \(\tilde{L}\gg1\), the period in particular approaches \(\pi E_T/2\).
Near zero temperature, the single-band CPR of a long junction may be approximated as a sawtooth-like function (see Fig. \ref{fig:I1D2})
\begin{equation}
    I^\text{1D}(0,\phi)=A\phi,\quad-\pi<\phi<\pi.\label{eq:sawtooth}
\end{equation}
Using this approximate form, one could obtain a total CPR that is qualitatively consistent with the exact results shown in Fig. \ref{fig:CPR1}.
Together with the relative phase shift $\delta\phi$ and the band DOS difference, this form of single-band CPR explains the JDE at small $h$ shown in Fig. \ref{fig:eta}, signaled by nonvanishing diode efficiency defined as
\begin{equation}
    \eta=\frac{I_{c+}-I_{c-}}{I_{c+}+I_{c-}}.
\end{equation}
The largest possible diode efficiency of such a ballistic long junction at zero temperature can hence be roughly estimated using Eq. (\ref{eq:sawtooth}) $|\eta|_\text{max}=3-2/(\tilde{\alpha}+1)-(\tilde{\alpha}+1)\leq3-2\sqrt{2}\approx 0.17$, close to the numerical result shown in Fig. \ref{fig:eta}.

When the magnetic field further increases and the superconductor enters the strong helical phase, the diode efficiency is greatly suppressed. This is because the current contribution from the lower DOS $\lambda=-$ band decreases significantly due to the emergence of BFS, as discussed earlier. The total current hence originates only from that of the higher DOS $\lambda=+$ band, resulting in a strong reduction of the Josephson current, as seen in Fig. \ref{fig:Ic}. Since a single band has negligible asymmetry of critical currents, the diode efficiency is also greatly reduced, as seen in Fig. \ref{fig:CPR2}.
For a short junction $\tilde{L}\ll 1$, as $\delta\phi^n$ is almost absent and the oscillation period of $\eta$ is significantly increased, the diode efficiency first reaches an extremum and then also gets suppressed at the entrance of strong helical phase, leading to behavior shown in Fig. \ref{fig:eta2}. This is qualitatively similar to the results obtained in \cite{DavydovaFu22} for \(\eta\) as a function of \(q\) instead of \(h\), but note that in our case \(q\) is determined self-consistently.

\begin{figure}[!h]
\centering    
\subfigure[]{    \centering
    \includegraphics[width=0.43\textwidth]{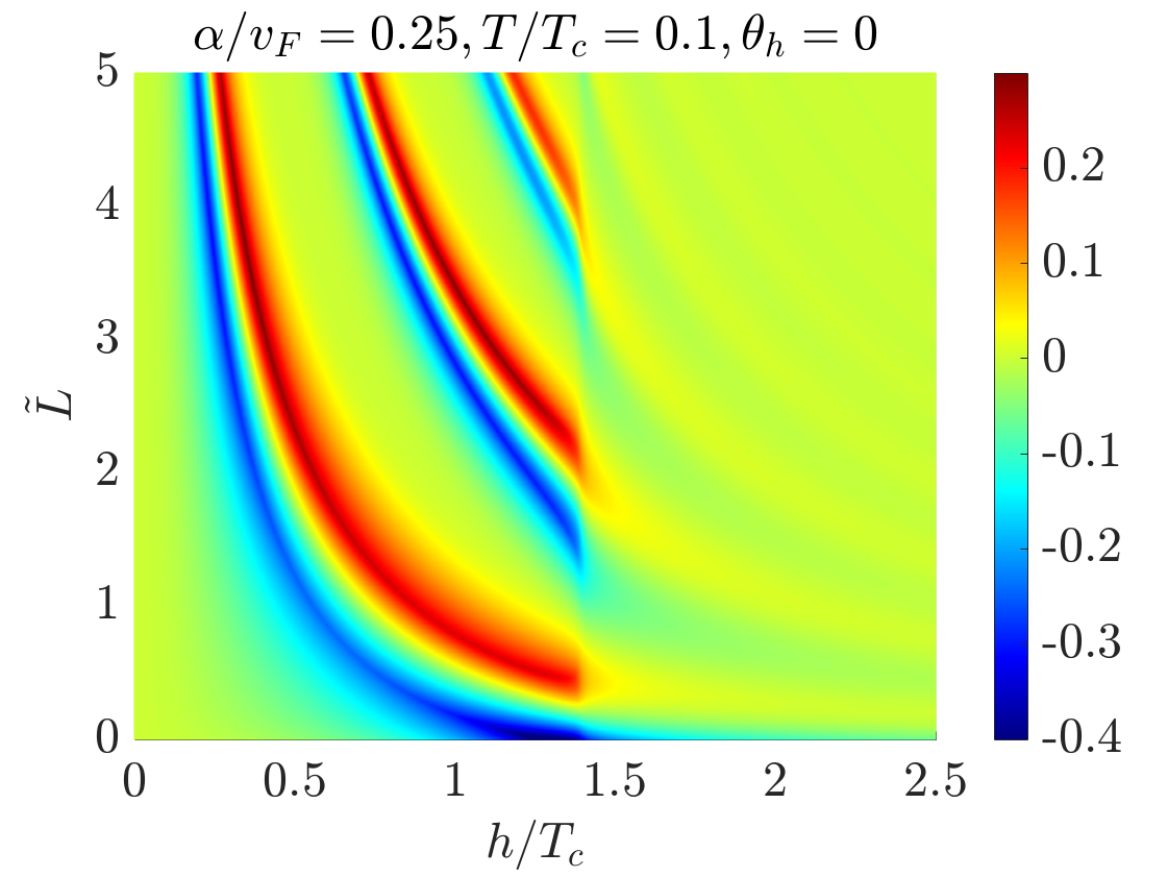}\label{fig:DiodeEff1}}
\subfigure[]{    \centering
    \includegraphics[width=0.43\textwidth]{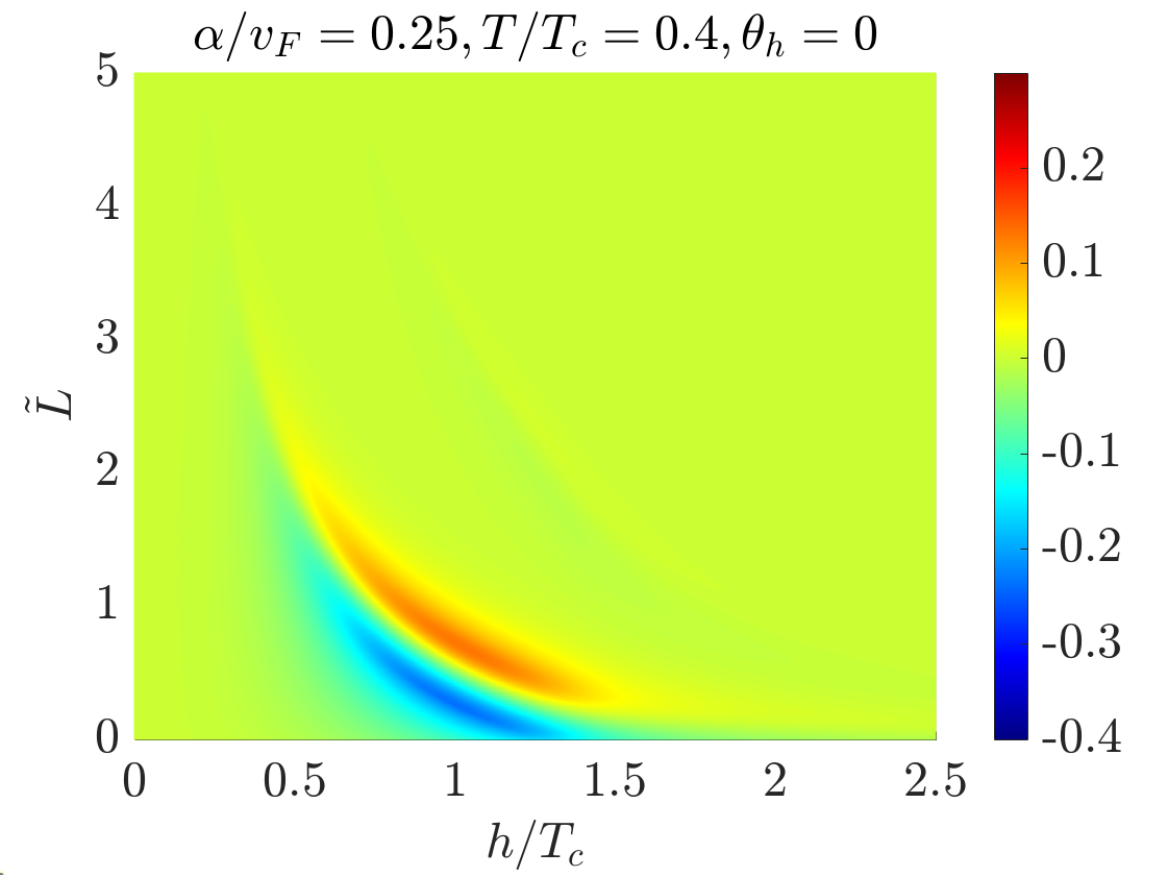}\label{fig:DiodeEff2}}
    \subfigure[]{    \centering
    \includegraphics[width=0.43\textwidth]{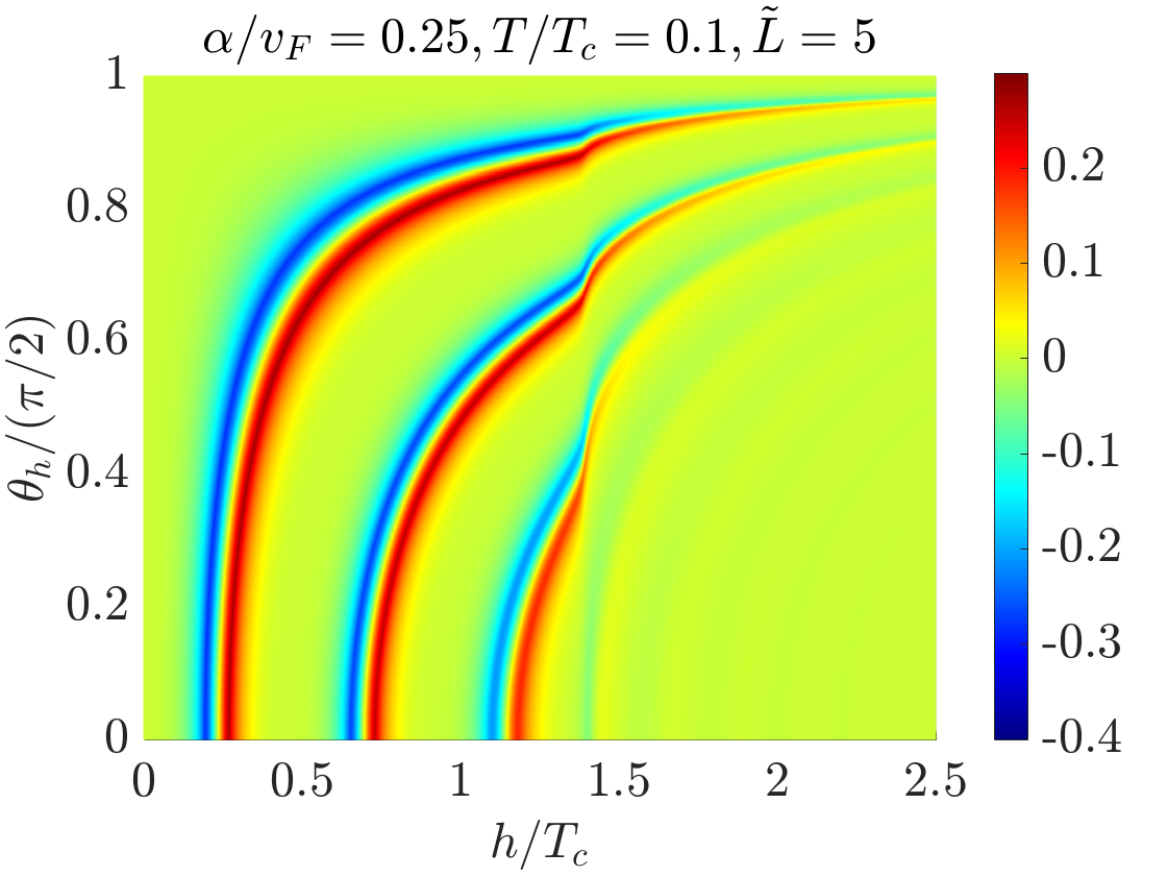}\label{fig:DiodeEff3}}
    \caption{The diode efficiency as a function of $\tilde{L}$, $\theta_h$ and $h$ at different temperatures. }
    \label{fig:DiodeEff}
\end{figure}

The oscillating characteristic of the diode efficiency discussed earlier is a general feature of such a nonreciprocal Josephson junction, shown in Fig. \ref{fig:DiodeEff1}. At low temperature, the diode efficiency oscillates with $\tilde{L}$ and $h$ due to the relative phase shift $\delta\phi$ between the bands, and then reduces significantly once $h/T_c\gtrsim1.44$, at which the superconductor enters the strong helical phase, in accordance with previous discussions. At higher temperatures, noticable JDE can only be seen in relatively short junctions $L\lesssim v_F/T$ as shown in Fig. \ref{fig:DiodeEff2}. This can be understood by noting that high temperature tends to destroy the higher harmonics in the single-band CPR, which is necessary for JDE.

\begin{figure}
\centering
        \subfigure[]{    \centering
    \includegraphics[width=0.45\textwidth]{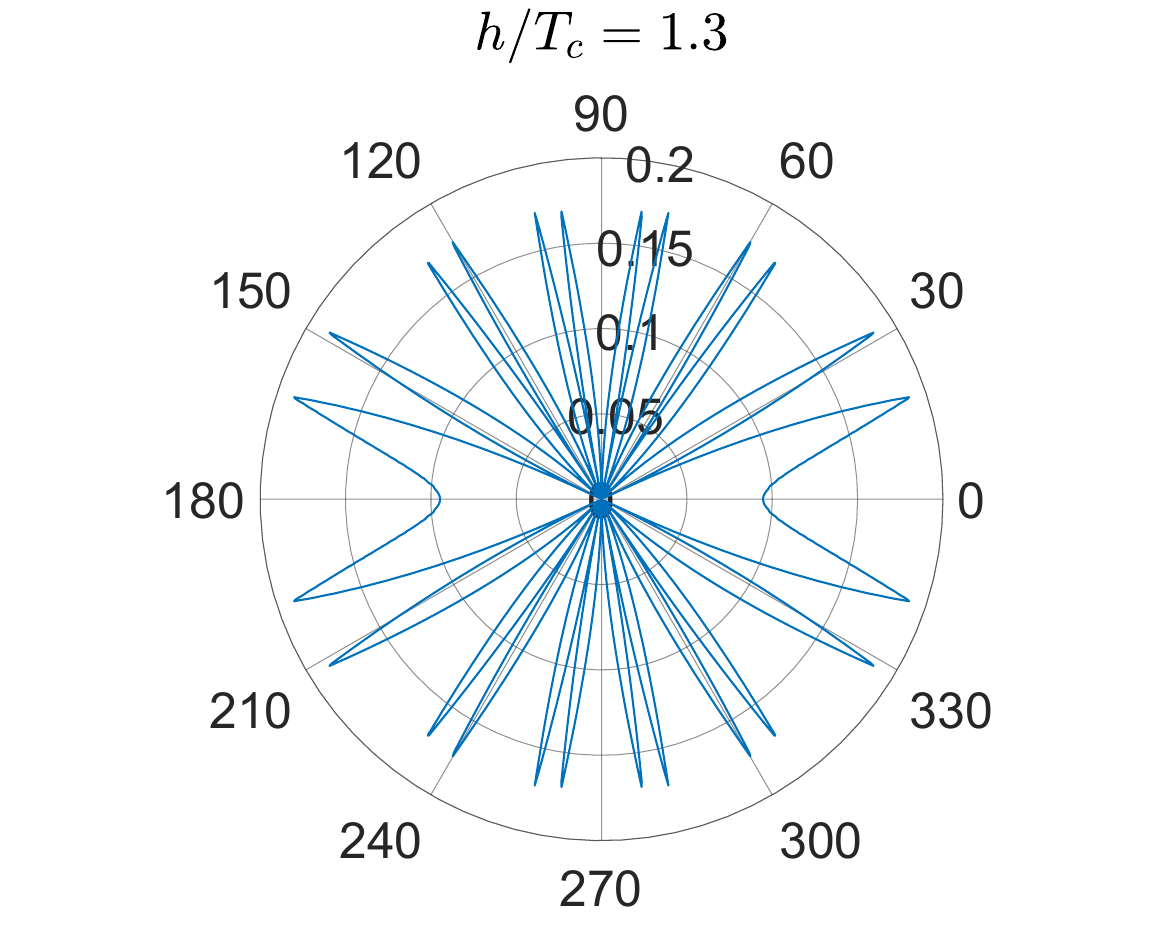}}
    \subfigure[]{    \centering
    \includegraphics[width=0.45\textwidth]{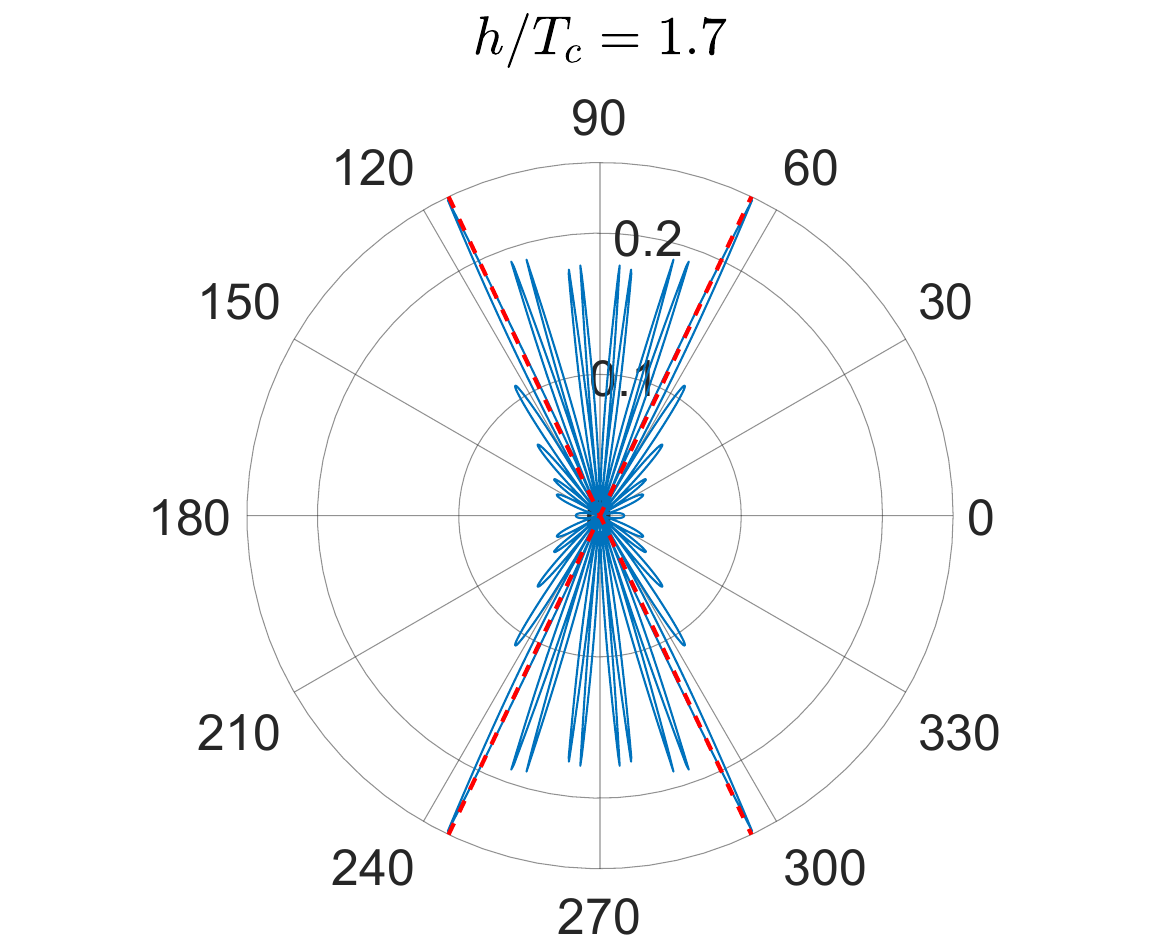}}
    \caption{The polar plot of $|\eta|$ as a function of $\theta_h$ at $T/T_c=0.01$, for (a) $h/T_c=1.3$ (weak helical phase) and (b) $h/T_c=1.7$ (strong helical phase). The red dashed line in (b) denotes the angle where BFS along $x$-direction appears.}
    \label{fig:etapolar}
\end{figure}

The discussion above assumed that $\mathbf{h}$ is perpendicular to the $x$-axis so that $\mathbf{q}$ (or \(\mathbf{Q}_\lambda\)) is parallel to the current. 
In Fig. \ref{fig:DiodeEff3} we consider how the diode efficiency evolves when the in-plane magnetic field direction is tilted away from the $y$-axis. Here we assume it is oriented at an angle $\theta_h$ with respect to the $y$-axis. 
From Eq. (\ref{Aeta}), one could see that the effects of $\theta_h$ is two-fold. First, it modifies the relative phase shift due to the normal metal to $\delta \phi^n=-4\tilde{h}\tilde{L}\cos \theta_h$, leading to diode efficiency oscillations as $\theta_h$ varies. Furthermore, it changes the superconducting gap experienced by the carriers in the quasi-1D normal metal--note that the superconducting gap only depends on the projection of $\mathbf{q}$ on x-axis, which is given by $|\tilde{\omega}|<1-|(\tilde{q}+2 \tilde{h})\cos \theta_h/2|$ according to Eq. (\ref{eq:LDOS}). Due to this $\theta$-dependent superconducting gap, or the anisotropy of BFS in momentum space, the Josephson current and diode efficiency exhibit strong anisotropy, especially in the strong helical phase, plotted in Fig. \ref{fig:etapolar}. For general $\theta_h$, the diode efficiency may still remain non-negligible when the superconductor is in the strong helical phase, in particular near $\theta_h\approx\pi/2$, as shown in Figs. \ref{fig:eta3}, \ref{fig:DiodeEff3} and  \ref{fig:etapolar}. For the special case \(\theta_h=\pi/2\), there is no JDE and $\eta$ vanishes exactly, due to the absence of relative phase shift between two bands.

\section{Discussion}\label{Sec:discussion}

In this work, we analyzed SDE and JDE in the helical superconductor and corresponding SNS junction with strong Rashba SOC in the clean limit. For the Josephson junction, we analytically solve the Eilenberger equation and derive the expression of the Andreev spectrum and current-phase relations (CPR) of each helical band. Due to the unequal band density of states as well as the existence of higher harmonics in the single-band CPR, the critical currents in opposite directions differ and lead to JDE. The in-plane magnetic field induces a finite Cooper pair momentum in the superconductor, as well as a relative phase shift between two bands, resulting in the oscillation of diode efficiency. At higher fields, the superconductor enters the strong helical phase and the Josephson current is significantly suppressed as only one of the helical bands dominates the transport due to the formation of the BFS. Similar phenomena have also been noticed in the case of a bulk superconductor, where SDE is strongly enhanced and changes sign near the transition, followed by eventual suppression due to the dominant DOS of one of the helical bands \cite{DaidoYanase22,Daido2022_2,IlicBergeret22}.  Interestingly, we also find that the bulk SDE may also approach its ideal limit, i.e. $|\eta|=1$, if the system is close to the CEP of the first-order phase transition, a previously overlooked possibility.

A key finding of our work is that both the diode efficiency and the critical current itself are strongly suppressed in the strong helical phase due to the presence of the BFS, especially at low temperatures. A similar suppression of the critical current had been noted in \cite{DolciniMeyerHouzet15}, but not its relation to the BFS or JDE.
These results demonstrate that long Josephson junctions are a promising platform for identifying helical SCs, particularly the transition into the strong helical phase marked by the suppression of the JDE. Moreover, we expect our results regarding the suppression of the Josephson current in a single-band scenario to hold qualitatively for any system with a BFS, regardless of the presence of JDE.
To the best of our knowledge little direct evidence for BFSs exists, and the only claim of direct observation of a BFS is based on quasi-particle interference (QPI) measurements of Bi\(_2\)Te\(_3\) proximitized to NbSe\(_2\) \cite{ZhuFu21}. On the other hand, a lot of indirect experimental evidence exists for BFSs in several materials, for example in uranium based SCs like U\(_{1-x}\)Th\(_x\)Be\(_{13}\) \cite{Zieve04, FisherTaillefer89, Heffner90}, UPt\(_3\) \cite{Schuberth92,JoyntTaillefer02}, URu\(_2\)Si\(_2\) \cite{Fisher90, Behnia92, Kasahara07, SchemmKapitulnik15}, and UPd\(_2\)Al\(_3\) \cite{Caspary93, ChiaoTaillefer97}; BFSs in the low-field SC states in UTe\(_2\) were also initially proposed
\cite{Ran19, Shick19}, but point nodes are favored by current consensus \cite{MetzAgterberg19}. Other materials like CeCoIn\(_5\) \cite{Movshovich01}, SrPtAs \cite{FischerGoryo15, BzdusekSigrist17, SumitaYanase19}, Sr\(_2\)RuO\(_4\) \cite{SuhAgterberg20}, Al-InAs heterostructures \cite{PhanShabaniSerbyn22}, as well as \(^3\)He \cite{LiuWilczek03, AuttiVolovik20}, also show evidence of BFSs. Fermi arcs measured in the pseudogap state of underdoped cuprates have also been interpreted as a Fermi surface in a PDW state \cite{BaruchOrgad08, BergFradkinKivelsonTranquada09, Agterberg20}, though alternative explanations exist. A lot of effort has been dedicated to measuring BFSs in FeSe\(_{1-x}\)Se\(_x\) \cite{Sato18, Hanaguri18, Nagashima22, Mizukami23, Matsuura23, Yu23, Nagashima24, Yu25}, with multiple theoretical studies
\cite{SettyHirschfeld20_I, CaoHirschfeld23, MikiHoshino24, IslamChubukov24, WuAgterberg24, CaoHirschfeld24}.

There is therefore an active demand for additional probes of BFSs that long Josephson junctions we considered in this work may fulfill.
The main challenge in detecting BFSs, as noted in the introduction, is that all the indirect evidence of BFSs is tied to their finite residual DOS, which can arise trivially due to disorder \cite{AG60}.
A similar issue is well-known in 1D SCs, where disorder-localized Andreev bound states can mimic the zero bias peak associated with topological Majorana bound states \cite{DasSarma21}.
Though the suppression of both the JDE and the Josephson currents are also a result of the residual DOS, a key distinction from the disorder-induced gapless SC is the prediction of a strong anisotropy of the Josephson current suppression with respect to current direction relative to the location of the BFS in momentum space, tied to the magnetic field in helical SCs. 
For helical SCs the anisotropy with respect to the field direction is expected to be two-fold symmetric. Interestingly, such a two-fold anisotropy has been observed in the 2D Ising SC 1H-NbSe\(_2\) \cite{ShafferBurnellFernandes20, HamillShafferPribiag21, ChoSchmalian22, HaimLevchenkoKhodas22}, though in the absence of an explicit junction geometry.

In principle, with a sufficiently narrow junction it may even be possible to partially image the BFS. It would be pertinent, therefore, to extend our study to 2D junctions and study the effects of the width of the junction.
A similar proposal was suggested using dissipative tunneling currents in NS junctions  that are expected to exhibit a similar anisotropy \cite{DavydovaFu24}, but a Josephson junctions would provide a more direct probe of the superconducting state. Ref. \cite{DavydovaFu24} additionally suggested a possible application in scanning tunneling microscopy (STM) experiments. Our proposal can similarly be applied in Josephson scanning tunneling microscopy (JSTM) \cite{SmakovMartinBalatsky01, KimuraAndoDynes08, RanderiaFeldmanYazdani16}, which has recently been used to detect finite momentum Cooper pairing \cite{HamidianSeamus16, LiuSeamus20, GuSeamus23}. Our results suggest that JSTM can thus potentially be used to also detect BFSs.

Given that disorder-induced gapless SC is the main effect that can mimic BFSs, it is also important to study how disorder affects Josephson junctions with BFSs, and especially whether the presence or absence of the anisotropy of the Josephson current suppression can indeed discriminate the two.
Several past studies have considered effects of disorder on systems with BFSs \cite{Fulde65, OhAgterberg21, BabkinSerbyn24, MikiHoshino24} and gapless SCs without BFSs \cite{Hauser67}, but the effect on JDE and the BFS-induced anisotropy is yet to be studied. We anticipate that disorder would suppress the diode efficiency in the strong SOC limit, as it generally suppresses higher harmonics in the CPR \cite{Golubov2004}, but that the anisotropy will survive if disorder is not too strong. Finally, we expect that BFSs will have strong signatures in thermal transport in Josephson junctions \cite{Mateos24}, which we also expect to be strongly anisotropic and possibly non-reciprocal. We leave these topics for future studies.

\begin{acknowledgements}
This work was financially supported by the National Science Foundation (NSF), Quantum Leap Challenge Institute for Hybrid Quantum Architectures and Networks Grant No. OMA-2016136 (Z. Z. and D. S.); NSF Grant No. DMR-2452658 (J. H. and A. L.) and H. I. Romnes Faculty Fellowship provided by the University of Wisconsin-Madison Office of the Vice Chancellor for Research and Graduate Education with funding from the Wisconsin Alumni Research Foundation.
\end{acknowledgements}

\bibliography{Ref}
\end{document}